\documentclass[traditabstract]{aa}
\usepackage{verbatim} 
\usepackage{amsmath}
\usepackage{amssymb}
\usepackage{graphicx}
\usepackage{booktabs}
\usepackage[authoryear]{natbib}
\usepackage[colorlinks=true,linkcolor=blue,urlcolor=blue,citecolor=blue]{hyperref}
\usepackage{xspace}
\usepackage{siunitx}
\usepackage[dvipsnames]{xcolor}

\makeatletter

\usepackage{txfonts}\usepackage{twoopt}
\bibpunct{(}{)}{;}{a}{}{,}
\usepackage{enumitem}

\newcommand{\msun}{{\rm M}_{\odot}}

\newcommand{\rsun}{{\rm R}_{\odot}}

\renewcommand{\vec}[1]{\boldsymbol{#1}}
\newcommand{\vectorunit}[1]{\boldsymbol{\hat{#1}}}

\newcommand{\eg}{e.g.\@\xspace}

\newcommand{\ie}{i.e.\@\xspace}
\newcommand{\mesa}{\texttt{MESA}\xspace}
\newcommand{\Cd}{\ensuremath{C_\mathrm{d}}\xspace}
\newcommand{\Ch}{\ensuremath{C_\mathrm{h}}\xspace}
\newcommand{\Rtau}{\ensuremath{R_{\tau = 10}}\xspace}
\newcommand{\recrad}{hydrogen-recombination radius\xspace}

\newcommand{\vb}[1]{#1}
\newcommand{\vbmath}[1]{#1}
\newcommand{\vbre}[1]{#1}
\newcommand{\vbmathre}[1]{#1}
\newcommand{\figref}[1]{Fig.~\ref{#1}}
\newcommand{\sectref}[1]{Sect.~\ref{#1}}
\newcommand{\eqtnref}[1]{Eq.~\eqref{#1}}
\newcommand{\tabref}[1]{Table~\ref{#1}}
\titlerunning{1D CE evolution}
\authorrunning{V.A.~Bronner}
\makeatother
\begin{document}
\title{Going from 3D to 1D: A one-dimensional approach to common-envelope evolution}
\author{V.A.~Bronner\inst{\ref{HITS}}\thanks{vincent.bronner@h-its.org}, F.R.N.~Schneider\inst{\ref{HITS},\ref{ARI}}, Ph.~Podsiadlowski\inst{\ref{Oxford}, \ref{HITS}}, F.K.~R\"{o}pke\inst{\ref{HITS},\ref{ITA}}}
\institute{Heidelberger Institut f\"{u}r Theoretische Studien, Schloss-Wolfsbrunnenweg 35, 69118 Heidelberg, Germany\label{HITS}
\and
Zentrum f\"{u}r Astronomie der Universit\"{a}t Heidelberg, Astronomisches Rechen-Institut, M\"{o}nchhofstr. 12-14, 69120 Heidelberg, Germany\label{ARI}
\and
University of Oxford, St Edmund Hall, Oxford OX1 4AR, UK\label{Oxford}
\and
Zentrum für Astronomie der Universit\"{a}t Heidelberg, Institut f\"{u}r Theoretische Astrophysik, Philosophenweg 12, 69120 Heidelberg, Germany\label{ITA}
}
\date{Received xxx / Accepted yyy}
\abstract{
    The common-envelope (CE) phase is a crucial stage in binary star evolution because the orbital separation can shrink drastically while ejecting the envelope of a giant star. Three-dimensional (3D) hydrodynamic simulations of CE evolution are indispensable to learning about the mechanisms that play a role during the CE phase. While these simulations offer great insight, they are computationally expensive. We propose a one-dimensional (1D) model to simulate the CE phase within the stellar evolution code \mesa by using a parametric drag force prescription for dynamical drag and adding the released orbital energy as heat into the envelope. We compute CE events of a \qty{0.97}{\msun} asymptotic giant-branch star and a point mass companion with mass ratios of 0.25, 0.50, and 0.75, and compare them to 3D simulations of the same setup. The 1D CE model contains two free parameters, which we demonstrate are both needed to fit the spiral-in behavior and the fraction of ejected envelope mass of the 1D method to the 3D simulations. For mass ratios of 0.25 and 0.50, we find good-fitting 1D simulations, while for a mass ratio of 0.75, we do not find a satisfactory fit to the 3D simulation as some of the assumptions in the 1D method are no longer valid. In all our simulations, we find that the released recombination energy is important to accelerate the envelope and drive the ejection.
}
\maketitle
%
%
%
%
\section{Introduction}\label{sec:introduction}

Common-envelope (CE) evolution, originally proposed by \citet{Paczynski1976}, is a phase in binary star evolution during which a giant star engulfs its more compact companion, leading to the formation of a shared envelope around the core of the giant star and the companion. Frictional forces cause the binary orbit to shrink. The released orbital energy expands the envelope, which can result in complete or partial envelope ejection while forming a close binary or a merger of the two stars if the envelope cannot be fully ejected (for reviews, see \citealt{Ivanova2013b}, \citealt{DeMarco2017} and \citealt{Roepke2023}). Today, it is believed that CE evolution is the main mechanism for converting wide binary star systems into close binaries \citep{Ivanova2013b}. Therefore, CE evolution plays an important, often essential role in understanding the formation channels of close binary systems: such as the CE channel for gravitational-wave sources (\citealt{Tutukov1993, Belczynski2002,Voss2003,Eldridge2016,Stevenson2017,Kruckow2018};\vb{\citealt{VignaGomez2018}};\citealt{Spera2019}; for a more complete summary of references see, \eg, \citealt{Mandel2022}), possibly the progenitors of SNe Ia \citep{Iben1984, Webbink1984, Whelan1973}, X-ray binaries \citep{Tauris2006}, cataclysmic variables \citep{Warner1995},  short-period hot subdwarfs \citep{Heber2009}, and even potentially gamma-ray burst sources  \citep{Fryer1998, Izzard2004, Detmers2008}.

Soon after CE evolution was proposed, one-dimensional (1D) simulations were carried out by \citet{Taam1978} and \citet{Meyer1979}. Using 1D hydrodynamic simulations, \citet{Podsiadlowski2001} found that the CE evolution can be divided into distinct phases. During the \emph{initial phase}, the co-rotation between the binary star and the envelope of the giant star is lost. Frictional forces then lead to the rapid spiral-in of the binary star during the \emph{plunge-in phase}, which is the shortest phase during CE evolution, happening on the dynamical timescale of the binary. As the CE expands, the spiral-in slows down, which might then lead to the \emph{self-regulated spiral-in phase}, its \emph{termination} and the \emph{post-CE evolution}, during which the changes of the orbital separation are small compared to the plunge in phase (see~\citet{Ivanova2013b} for the definitions of the final stages). For the following discussion in this work, we follow the approach of \citet{Roepke2023} and combine all stages after the plunge-in phase to the \emph{post-plunge-in phase}.

Three dimensional (3D) (magneto-)hydrodynamic simulations of CE events (\citealt{Ricker2008, Ricker2012, Passy2012, Nandez2014, Ohlmann2016a, Ohlmann2016b, Iaconi2017, Chamandy2018, Prust2019, Reichardt2019, Sand2020, Ondratschek2022, Lau2022a, Lau2022b, Moreno2022}) are currently the best tool to study the physical mechanisms that cause the spiral-in of the binary and the ejection of the envelope. These simulations are computationally expensive and easily need more than \qty{e5} core-hours. They allow us to study CE events in a case-by-case study, but they are infeasible when studying larger populations of systems undergoing CE evolution.

To predict the outcome of a CE event, an energy formalism was introduced \citep{Webbink1984, Livio1988}. The released orbital energy is compared to the binding energy of the envelope, where the CE efficiency $\alpha_\mathrm{CE}$ determines the fraction of the orbital energy that is used to unbind the envelope. Classically, an efficiency $\alpha_\mathrm{CE} \leq 1$ is assumed to account for energy conservation. However, energy sources other than the orbital energy might be available, \eg, the ionization energy from the recombination of hydrogen and helium \citep{Han1995, Ivanova2015}, accretion on the companion \citep{Chamandy2018}, the formation of jets \citep{Shiber2019}, and dust formation \citep{Glanz2018, Iaconi2020, Reichardt2020}. These extra energy sources allow physical scenarios with $\alpha_\mathrm{CE} > 1$. Therefore, the energy formalism cannot be used directly by itself to predict the outcome of a CE event, since the efficiency parameter $\alpha_\mathrm{CE}$ varies from system to system \citep{Iaconi2019}. Recent efforts have been made by \citet{Marchant2021} and \citet{Hirai2022} to improve on the energy formalism.

Even though 3D simulations are constantly improving and more computational power is available, 1D simulations of the CE phase remain a useful tool \citep[cf.][]{Ivanova2016, Clayton2017, Fragos2019, Trani2022}. The comparatively low computational costs of 1D simulations compared to 3D simulations open the opportunity to study a larger number of CE systems and explore the possible parameter space \citep{Ivanova2013b}. This makes 1D simulations more versatile than 3D simulations. \vb{In addition to 1D simulations, semi-analytic models calibrated on 3D simulation \citep[\eg,][]{Trani2022} have similar advantages.}

As outlined above, it is still a difficult task to predict the outcome of CE events, because the input physics of 1D methods are incomplete, prohibiting accurate simulations of CE evolution, while 3D simulations are computationally too expensive. We introduce a new 1D method, to efficiently compute CE events. The assumptions made in 1D models, such as spherical symmetry and energy deposition in shells, do not capture the physical processes involved in CE events. To compensate for this, we introduce free parameters, which we calibrate on the characteristics (\eg, orbital separation, the mass of ejected envelope) of 3D simulations. We test to see how many parameters are needed, such that the 1D simulation can reproduce the characteristics of 3D simulations. Our method needs to be calibrated on 3D simulations, to obtain the most accurate results while still retaining the computational advantages of a 1D simulation. If possible, the calibrated method can be used to simulate CE events and predict their outcome.

This paper is structured as follows. In \sectref{sec:methods}, we describe our new 1D CE method. Then, we apply this method to simulate CE events of a \qty{0.97}{\msun} asymptotic giant (AGB) star and point-mass companions. Detailed results of the simulation with a mass ratio of 0.25 are shown in \sectref{sec:results_q0.25}, before presenting the results for simulations with mass ratios of 0.50 and 0.75 in \sectref{sec:results_higher_q}. Finally, we discuss our proposed model and the results in \sectref{sec:discussion}, and conclude in \sectref{sec:conclusion}.

%
%
\section{Methods}\label{sec:methods}
We model the 3D CE simulation of \citet{Sand2020} in the 1D stellar-evolution \mesa revision 12778 \citep{Paxton2011, Paxton2013, Paxton2015, Paxton2018, Paxton2019}. We describe how to obtain the same pre-CE giant star as \citet{Sand2020} in \sectref{sec:initial_model} and our general hydro setup in \sectref{sec:hydro_MESA}. For numerical reasons, we first perform a relaxation run of the pre-CE model, which we explain in \sectref{sec:relaxation}. Frictional forces are hindering the motion of the companion inside the envelope and cause the orbit to decay. In \sectref{sec:orbital_evolution}, we introduce a parametric prescription of this drag force and show how the drag force modifies the equations of motion that we integrate to obtain the binary orbit. The mass of the companion modifies the gravitational potential of the giant star, which we describe in \sectref{sec:mod_G}. In \sectref{sec:CE_heating}, we show how we model the back reaction of the companion on the envelope by artificially heating the envelope layers around the companion. The unbound layers at the surface of the CE are removed from the simulation, as explained in \sectref{sec:mass_loss}. Within this model, we use two calibration parameters; one parameter determines the strength of the drag force and the other sets the size of the artificially heated zone. In \sectref{sec:compare_3D} we show how we calibrate these parameters by comparing our 1D simulations to 3D hydrodynamic simulations of CE events of \citet{Sand2020}.

\subsection{Initial model}\label{sec:initial_model}
We evolve an initially \qty{1.2}{\msun} star with metallicity $Z = 0.02$ from the zero-age main sequence to the ABG phase and stop the evolution once the model reaches a mass of \qty{0.97}{\msun}. For this run, we use the default \mesa settings, except for changing the wind-loss parameters in the Reimers prescription to $\eta_\mathrm{R} = 0.5$ \citep{Reimers1975} and in the Bl\"ocker prescription to $\eta_\mathrm{B} = 0.1$ \citep{Bloecker1995}. Additionally, we disable the \mesa `gold tolerances' to evolve the model through the helium flash, and we use a mixing-length parameter of $\alpha_\mathrm{MLT} = 2$.

\begin{figure}
    \resizebox{\hsize}{!}{\includegraphics{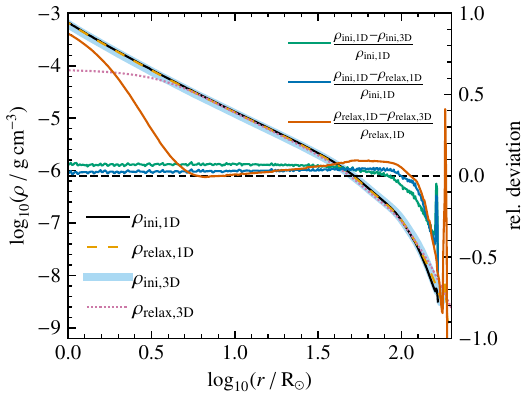}}
    \caption{The initial density profile $\rho_\mathrm{ini, 1D}$ of the AGB star in comparison to the initial density profile $\rho_\mathrm{ini,3D}$ of the AGB star in \citet{Sand2020}, as well as the \vb{1D and 3D density profiles} after the relaxation run $\rho_\mathrm{relax,1D}$ \vb{and} $\vbmath{\rho_\mathrm{relax,3D}}$. Relative deviations between the density profiles are shown along the secondary ordinate.}
    \label{fig:compare_ini_rho}
\end{figure}

Our initial model of the ABG star has a radius of \qty{166}{\rsun}. This is similar to the radius of \qty{172}{\rsun} for the initial model of \citet{Sand2020}. The density in the envelope of our initial model is ${\sim}\, 5 - 10\, \%$ larger compared to the initial model for the 3D simulations (\figref{fig:compare_ini_rho}), with deviations of up to $45 \, \%$ close to the surface. We attribute this to the older \mesa revision 7624 used to construct the initial pre-CE model in \citet{Sand2020}, which makes it difficult to reproduce the exact same model.

\subsection{Hydrodynamic setup in \mesa}\label{sec:hydro_MESA}
We simulate the CE evolution in \mesa using its single star module \texttt{star} together with an artificial heating source to account for the presence of a companion star. We switch to the implicit hydrodynamics in \mesa by using an HLLC solver \citep{Paxton2018}. With implicit hydrodynamics, we can study the dynamical evolution of the CE and also capture potential shocks. To better resolve shocks near the surface of the envelope, we impose a vanishing compression gradient for the surface boundary condition that sets the surface density \citep{Paxton2015, Grott2005}. For the other surface boundary condition, here the temperature, we use a black body. 

We use an adaptive mesh refinement in the velocity, optimized for hydrodynamic simulations in \mesa \citep{Paxton2018}, with uniform spacing in $\log\,r$ for the cells. The number of cells is chosen to be similar to the number of cells in the initial hydrostatic model. As this mesh refinement does not support rotation of the star, we simulate the CE without accounting for rotation in the stellar structure equations.

The hydrodynamic equations with \mesa's HLLC scheme are solved implicitly in time. Thus, the time-steps of our simulations are not limited by the Courant-Friedrichs-Lewy (CFL) criterion \citep{Courant1928, Paxton2018}. However, to resolve shock waves that travel through the envelope, we limit the time-steps to \qty{45}{\%} of the maximum time-step given by the CFL criterion. Furthermore, we find that limiting the time-steps to $ < \qty{e-4}{yr}$ helps with the numerical stability of the simulations. All the 1D CE simulations are run for as long as possible and stop for numerical reasons, \eg, too short time-steps.

\subsection{Relaxation run}\label{sec:relaxation}
Before starting the CE simulation, we perform a relaxation, where we switch from the standard hydrostatic solver to the implicit hydrodynamic solver and change the surface boundary conditions. The relaxation run lasts for \qty{2}{yr} or about 15 dynamical timescales\footnote{We start the relaxation run with an initial time-step of \qty{e-7}{yr} and set the maximum time-step during the relaxation run to \qty{e-3}{yr}.}. During the final \qty{0.2}{yr} of the relaxation run, the maximum time-step is linearly decreased to a value of \qty{e-7}{yr}. We find that this helps to start the CE simulation from the relaxed model without numerical artifacts.

During the relaxation run, the envelope expands to \qty{189}{\rsun}. The density profile of the envelope of the AGB star changes by less than \qty{5}{\%} during the relaxation run (\figref{fig:compare_ini_rho}). At the end of the relaxation run, the density close to the surface of the star decreased by \qty{60}{\%}. This shows that mostly the surface layers of the star expand during the relaxation run, leaving the envelope structure unchanged. \vb{\figref{fig:compare_ini_rho} also compares the density profile after the relaxation run to the relaxed density profile of \citet{Sand2020}. Throughout most parts of the envelope, the density profiles deviate by less than \qty{10}{\%}, except for larger deviations towards to surface. The deviations towards smaller radii originate from the cut-out core in the 3D simulation.}

\subsection{Orbital evolution}\label{sec:orbital_evolution}
We integrate the equations of motion of the classical two-body problem consisting of an extended sphere at location $\vec{x}_1$, representing the giant star, and a point-mass particle at location $\vec{x}_2$, representing the companion, to compute the orbital evolution of the binary star. In general, quantities with index 1 refer to the giant star, while index 2 refers to the companion. As the companion is revolving within the giant star during the CE event, only the mass of the giant star enclosed by the orbit  $M_{1,a_\mathrm{orb}}$ needs to be considered in the equations of motion, 
\begin{align}
    \vec{\ddot{x}}_1 &= \dfrac{G M_2}{a_\mathrm{orb}^2}\vectorunit{a}_\mathrm{orb} \label{eq:eqs-of-motion-1} \\
    \vec{\ddot{x}}_2 &= -\dfrac{G M_{1,a_\mathrm{orb}}}{a_\mathrm{orb}^2}\vectorunit{a}_\mathrm{orb} - \dfrac{F_\mathrm{d}}{M_2}\vectorunit{v}_\mathrm{rel} \label{eq:eqs-of-motion-2}
\end{align}
 where the orbital separation, \ie, the distance from the center of the giant star to the point-mass companion, is given by $\vec{a}_\mathrm{orb} = \vec{x}_2 - \vec{x}_1$. The mass of the companion is given by $M_2$ and $G$ is the gravitational constant. Vector quantities in non-bold notation denote the absolute value of that quantity and $\vectorunit{x}$ is a unit vector in the direction of $\vec{x}$. Because the companion moves within the envelope of the giant star, it is subject to a drag force $\vec{F}_\mathrm{d} = -F_\mathrm{d} \vectorunit{v}_\mathrm{rel}$ caused by dynamical friction, opposing the relative velocity $\vec{v}_\mathrm{rel}$ of the companion to the envelope.
 
 We use the semi-analytic formalism of \citet{Kim2010} and \citet{Kim2007} to calculate the drag force. \citet{Kim2007} extended the classical model of dynamical friction of \citet{Ostriker1999} to perturbers moving in circular orbits with radius $r$ through a homogeneous background medium of density $\rho$. \citet{Kim2010} then extended the model further to account for possible non-linear effects arising from the circular motion of the perturber. They define the two dimensionless parameters
 \begin{equation} 
    \mathcal{B} = \frac{G M_2}{c_\mathrm{s}^2 r}, \qquad \eta_\mathcal{B} = \frac{\mathcal{B}}{\mathcal{M}^2 - 1},
\end{equation}
with $M_2$ the mass of the perturber, \ie, the companion in our case, $c_\mathrm{s}$ the sound speed, and $\mathcal{M} = v_\mathrm{rel}/c_\mathrm{s}$ the Mach number. They find a correction term $f_\rho$ for the density to account for non-linear effects,
\begin{equation}
    \rho_\mathrm{nl} = f_\rho \rho = \left(1 + \frac{0.46\, \mathcal{B}^{1.1}}{{(\mathcal{M}^2 - 1)}^{0.11}}\right) \rho.
\end{equation}
Together, the drag force is then given by
\begin{equation}\label{eq:F_d}
    F_\mathrm{d} = 
    \begin{cases}
        \vbmath{\Cd} \dfrac{4 \pi \rho_\mathrm{nl}{(G M_2)}^2}{v_\mathrm{rel}^2} \left( \dfrac{0.7}{\eta_\mathcal{B}^{0.5}} \right) \quad & \mathrm{for}\, \eta_\mathcal{B} > 0.1 \wedge  \mathcal{M} > 1.01, \\[1em]
        \vbmath{\Cd} \dfrac{4 \pi \rho {(G M_2)}^2}{v_\mathrm{rel}^2} I & \mathrm{else},
    \end{cases}
\end{equation}
where the condition $\eta_\mathrm{B} > 0.1$ and $\mathcal{M}>1.01$ corresponds to a regime in which non-linear effects are present that modify the drag force, as opposed to the classical linear regime for $\eta_\mathcal{B} < 0.1$ or $\mathcal{M}<1.01$. \vb{The drag-force parameter \Cd is introduced as a free parameter in our model and modifies the strength of the drag force}. We modified the original condition by \citet{Kim2010} for the non-linear regime from $\mathcal{M} > 1$ to $\mathcal{M} > 1.01$ in \eqtnref{eq:F_d} to avoid divergence issues in the definitions of $\eta_\mathcal{B}$ and $\rho_\mathrm{nl}$. According to \citet{Kim2007}, the Coulomb logarithm $I$ in the linear regime is given by
\begin{equation}
    I = \begin{cases}
        0.7706 \ln \left(\dfrac{1 + \mathcal{M}}{1.0004 - 0.9185\, \mathcal{M}} \right) - 1.473\, \mathcal{M} \quad \mathrm{for}\, \mathcal{M} < 1.0, \\[1em]
        \ln \left(330 \dfrac{r}{r_\mathrm{min}} \dfrac{(\mathcal{M} - 0.71)^{5.72}}{\mathcal{M}^{9.58}}\right) \quad \mathrm{for}\, 1.0 \leq \mathcal{M} < 4.4, \\[1em]
        \ln \left( \dfrac{r / r_\mathrm{min}}{0.11\,\mathcal{M} + 1.65}\right) \quad  \mathrm{for}\, 4.4 \leq \mathcal{M}.
    \end{cases}
\end{equation}
We choose $r_\mathrm{min} = 3.1 \,\rsun$, a measure of the size of the companion, which is the same value as the softening-length of the gravitational potential of the companion in the simulations of \citet{Sand2020}.

\vb{
    The initial orbital separation $a_\mathrm{orb,ini}$ in \citet{Sand2020} is chosen such that the giant star overfills its Roche lobe, initiating the CE event. In particular, the companion is initially outside of the giant star, \ie, $a_\mathrm{orb,ini} > R_1$. This setup of the initial orbit would not lead to a CE event in our 1D model on a dynamical timescale, because the companion does not exert a tidal force on the giant star. Additionally, we assume a vanishing density around the giant star which leads to a vanishing drag force and no spiral-in behavior. Therefore, to initiate the CE event, we choose $a_\mathrm{orb,ini}=\qty{160}{\rsun}$, \ie, we place the companion well inside the giant's envelope of \qty{166}{\rsun} before the relaxation. The initial orbit is set up to be circular, following the setup in \citet{Sand2020}.
}

\vb{
    The giant star in the simulations of \citet{Sand2020} is initially set up to rotate rigidly at \qty{95}{\%} of the initial orbital frequency with the angular velocity vector $\vec{\Omega}$ pointing perpendicular to the orbital plane. We use the same $\vec{\Omega}$ for our 1D simulations as in \citet{Sand2020}, which means in particular that $\vec{\Omega}$ varies with mass ratio $q=M_2/M_1$.
}

\vb{
    The relative velocity of the companion to the envelope of the giant star is given by
    \begin{equation}\label{eq:v_rel}
        \vec{v}_\mathrm{rel} = \vec{\dot{x}}_2 - \vec{\dot{x}}_1 - \left(\vec{\Omega} \times \vec{a}_\mathrm{orb} + v_r(a_\mathrm{orb}) \vectorunit{a}_\mathrm{orb}\right),
    \end{equation}
    where $v_r(a_\mathrm{orb})$ is the (radial) expansion velocity of the CE at the location of the companion. While the giant star in the \mesa simulation does not rotate (\sectref{sec:hydro_MESA}), a rigidly rotating envelope with a constant angular velocity $\vec{\Omega}$ in time is assumed for calculating the relative velocity. The magnitude of the relative velocity enters the calculation of the drag force via \eqtnref{eq:F_d}.
}

The equations of motion (Eqs. \ref{eq:eqs-of-motion-1} and \ref{eq:eqs-of-motion-2}) are integrated in parallel with the \mesa simulation using a fifth-order explicit Runge-Kutta method with an embedded fourth-order error estimation \citep{Cash1990}. Relative and absolute error tolerances are set to $10^{-10}$, and the maximum integration time-step is set so that at least ten orbit-integration steps are taken during one \mesa time-step.

\subsection{Modified gravitational potential}\label{sec:mod_G}
For the simulations of the CE evolution with our 1D model, we use the single-star module \texttt{star} of \mesa to model the giant star. However, because of the mass of the companion, which is revolving inside the envelope of the giant star, the gravitational potential can no longer be approximated by that of a single star. To account for the change in the gravitational potential, we modify the gravitational constant $G$ to a radius-dependent gravitational constant $\tilde{G}(r)$ with
\begin{equation}\label{eq:mod_G}
    \tilde{G}(r) =
    \begin{cases}
        G \qquad & \mathrm{for\, } r < a_\mathrm{orb}, \\ 
        G\left(1 + \dfrac{M_2}{M_{1,r}}\right) \qquad & \mathrm{for\,} r \geq  a_\mathrm{orb},
    \end{cases}
\end{equation}
where $M_{1,r}$ is the mass of the giant star within radius $r$ \citep{Podsiadlowski1992}. This ensures that the layers outside the orbit are bound more strongly to the system because of the additional mass of the companion. Layers within the orbit are not affected. Changing the gravitational constant modifies not only the binding energy of the outer layers but also the entire envelope profile, \eg, density and pressure.

\subsection{CE heating}\label{sec:CE_heating}
The drag force acting on the companion inside the CE dissipates the orbital energy of the binary system, causing the orbital separation to shrink. The rate at which orbital energy is dissipated is given by $\dot{E}_\mathrm{dis} = \vec{F}_\mathrm{d} \cdot \vec{v}_\mathrm{rel}$. We add the dissipated energy from the binary orbit by artificially increasing the internal energy of the envelope, \ie, heating with the same rate as the orbital energy is lost, \ie, $\dot{E}_\mathrm{heat} = -\dot{E}_\mathrm{dis}$.

The range over which energy is injected is approximated by the accretion radius of the Bondi-Lyttleton-Hoyle accretion model \citep{Hoyle1941,Bondi1944}. The accretion radius profile $R_\mathrm{a}(r)$ inside the envelope is given by
\begin{equation}\label{eq:R_a}
    R_\mathrm{a}(r) = \frac{2 \, G M_2}{{\Delta v(r)}^2 + {c_\mathrm{s}(r)}^2},
\end{equation}
where $\Delta v(r)$ is the relative velocity of the spherical shell at coordinate $r$ with respect to the companion, and $c_\mathrm{s}(r)$ is the sound speed. The relative velocity is given by 
\begin{equation}
    \Delta v(r)^2 = \Delta v_r(r)^2 + \Delta v_\varphi(r)^2, 
\end{equation}
where
\begin{align}
    \Delta v_r(r) &= v_{\mathrm{orb},r} - v_r(r) \quad \text{and} \\
    \Delta v_\varphi(r) &= v_{\mathrm{orb},\varphi} - \Omega r.
\end{align}
In the above equations, $\vec{v}_\mathrm{orb} = \vec{\dot{x}}_2 - \vec{\dot{x}}_\mathrm{\vb{1}}$ is the orbital velocity and $v_r(r)$ is the (radial) expansion velocity of the envelope. The quantities $v_{\mathrm{orb},r}$ and $v_{\mathrm{orb},\varphi}$ are the projections of the orbital velocity vector along the $r$-direction and the $\varphi$-direction.

The accretion radius defines the area of influence of the companion in the giant's envelope. We choose to heat all the envelope layers with a radial coordinate between $r^\mathrm{heat}_{\min}$ and $r^\mathrm{heat}_{\max}$, which are determined by solving 

\begin{align}\label{eq:r_heat_min}
    r^\mathrm{heat}_{\min} &= a_\mathrm{orb} - \vbmath{\Ch} R_\mathrm{a}(r^\mathrm{heat}_{\min}), \\ \label{eq:r_heat_max}
    r^\mathrm{heat}_{\max} &= a_\mathrm{orb} + \vbmath{\Ch} R_\mathrm{a}(r^\mathrm{heat}_{\max}).
\end{align}
\vb{The heating parameter \Ch is introduced as a free parameter in our models and determines the extent of the heating zone by modifying the upper and lower boundary for the heating zone via Eqs.~\eqref{eq:r_heat_min} and \eqref{eq:r_heat_max}.} If there are multiple solutions\footnote{The accretion radius $R_\mathrm{a}(r)$ is not a monotonically increasing function, allowing multiple solutions for Eqs. (\ref{eq:r_heat_min}) and (\ref{eq:r_heat_max}).} for Eqs. (\ref{eq:r_heat_min}) and (\ref{eq:r_heat_max}), we use the largest one for $r^\mathrm{heat}_{\min}$ and the smallest one for $r^\mathrm{heat}_{\max}$, \ie, the solutions closest to $a_\mathrm{orb}$. This definition ensures that for all layers with $r^\mathrm{heat}_{\min} \leq r \leq r^\mathrm{heat}_{\max}$ the radial separation from the companion is smaller than the local accretion radius, \ie, these layers are gravitationally deflected/focused by the companion. The heating rate $\dot{\epsilon}_\mathrm{heat}$ then follows
\begin{equation}\label{eq:edot_heat}
    \dot{\epsilon}_\mathrm{heat} {\sim}\, \begin{cases}
        \exp\left[-\left(\dfrac{a_\mathrm{orb} - r}{a_\mathrm{orb} - r^\mathrm{heat}_{\min}}\right)^2 \right] \quad & \mathrm{for}\,  r^\mathrm{heat}_{\min} \leq r < a_\mathrm{orb}, \\[1em]
        \exp\left[-\left(\dfrac{a_\mathrm{orb} - r}{a_\mathrm{orb} - r^\mathrm{heat}_{\max}}\right)^2 \right] & \mathrm{for}\, a_\mathrm{orb} \leq r \leq r^\mathrm{heat}_{\max},\\[1em]
        0 & \mathrm{else},
    \end{cases}
\end{equation}
which is normalized so that the total heating rate matches the dissipation rate of the orbital energy, \ie, $\int \dot{\epsilon}_\mathrm{heat}\, \mathrm{d}m = \dot{E}_\mathrm{heat}$. In the case where $r^\mathrm{heat}_{\max}$ is undefined by \eqtnref{eq:r_heat_max} because $r \neq a_\mathrm{orb} + R_\mathrm{a}(r)$ for $a_\mathrm{orb} \leq r \leq R_1$, we choose $r^\mathrm{heat}_{\max} = R_1$. The lower heating limit $r^\mathrm{heat}_{\min}$ is always well defined because the sound speed increases by orders of magnitude toward the giant's core, causing the accretion radius to decrease.

\subsection{Dynamical envelope ejection}\label{sec:mass_loss}
The heating perturbs the hydrostatic equilibrium of the envelope, causing it to expand on the dynamical timescale. Envelope layers exceeding the local escape velocity $v_\mathrm{esc}$ are formally unbound. The local escape velocity is given by
\begin{equation}
    v_\mathrm{esc}(r) =  \sqrt{\dfrac{2 \, \tilde{G}(r) M_{1,r}}{r}},
\end{equation}
where $\tilde{G}(r)$ is the modified gravitational constant defined by \eqtnref{eq:mod_G}. We remove all surface layers with $v_r > v_\mathrm{esc}$ in our CE simulations to avoid numerical difficulties in the unbound layers. If a continuous layer of unbound mass\footnote{\vb{We only use the radial velocity $v_r$ in the criterion for envelope ejection, \ie, we neglect the kinetic energy in the rotation of the envelope. We find that excluding the kinetic energy in rotation changes the fraction of the ejected envelope mass in the 3D simulations of \citet{Sand2020} by less than \qty{2}{\%} after the end of the dynamical plunge-in phase.}} $M_\mathrm{unbound}$, \ie, $v_r > v_\mathrm{esc}$ for all shells in this layer, reaches the outer boundary of the simulation domain, we remove the layer exponentially by adopting a mass-loss rate 
\begin{equation}\label{eq:M_dot}
    \dot{M} = - \frac{M_\mathrm{unbound}}{\tau_{\dot{M}}},
\end{equation}
where $\tau_{\dot{M}}$ is the mass-loss timescale. In all our simulations, we choose $\tau_{\dot{M}} = \qty{0.01}{yr}$, which is shorter than both the dynamical timescale and the thermal timescale of the envelope. The applied mass-loss prescription is the same as that in \citet{Clayton2017}.

\subsection{Comparison to 3D CE simulations}\label{sec:compare_3D}
We compare and fit our 1D results to those of \citet{Sand2020}. In particular, we compare the time evolution of the orbital separation $a_\mathrm{orb}(t)$, from hereon called the spiral-in curve, and the mass fraction of the ejected envelope $f_\mathrm{ej}$, 
\begin{equation}\label{eq:f_ej}
    f_\mathrm{ej} =  \frac{M_\mathrm{env,ini} - M_\mathrm{env}}{M_\mathrm{env,ini}},
\end{equation}
where $M_\mathrm{env} =  M_1 - M_\mathrm{1,core}$ is the envelope mass, $M_1$ is the mass of the giant star and $M_\mathrm{1,core} = \qty{0.545}{\msun}$\footnote{The value of $M_\mathrm{1,core}$ is taken from \citet{Sand2020} to ensure a meaningful comparison.}. In contrast to this, \citet{Sand2020} defined the total unbound mass, and hence the mass fraction of the ejected envelope, by summing all cells with positive total energy, \ie, the sum of the potential and kinetic energy. To better match this quantity, we define a second envelope-ejection fraction,
\begin{align}\label{eq:f_ej_all}
    f_\mathrm{ej,all} = \frac{M_\mathrm{env,ini} - M_\mathrm{env,bound}}{M_\mathrm{env,ini}} \quad \mathrm{with}\\
    M_\mathrm{env,bound} = M_\mathrm{env} - M_\mathrm{unbound},
\end{align}
where $M_\mathrm{unbound}$ is the total mass of all layers in the envelope with $v_r > v_\mathrm{esc}$. The difference between $f_\mathrm{ej}$ and $f_\mathrm{ej,all}$ is that $f_\mathrm{ej,all}$ also includes the layers with $v_r > v_\mathrm{esc}$ within the envelope.

\vb{We incorporate two free parameters in our model. The drag-force parameters \Cd modifies the strength of the drag force via \eqtnref{eq:F_d} and the heating parameter \Ch modifies the extent of the heated layers via Eqs~\eqref{eq:r_heat_min} and \eqref{eq:r_heat_max}.} Both parameters are used to fit the spiral-in curves and the mass fraction of the ejected envelope material from the 1D CE simulations to the results of the corresponding 3D simulations in \citet{Sand2020}.

The initial orbital separations of our 1D simulations deviate from the initial separations used by \citet{Sand2020} (see~\sectref{sec:orbital_evolution}). Therefore, the spiral-in curve of the 3D simulation is shifted by $\Delta t$ with respect to the spiral-in curve of the 1D simulation. The optimal time shift $\Delta t_\mathrm{min}$ for fixed \Cd and \Ch is found by minimizing the mean relative deviation (MRD) between the spiral-in curves,
\begin{equation}\label{eq:MRD}
    \mathrm{MRD}(\Delta t) = \frac{1}{T} \sum_i \frac{\left|a_\mathrm{orb}^\mathrm{1D}(t_i) - a_\mathrm{orb}^\mathrm{3D}(t_i - \Delta t)\right|}{a_\mathrm{orb}^\mathrm{1D}(t_i)} \delta t_i,
\end{equation}
where $T$ is the total simulated time, $\delta t_i$ is the time-step and $a_\mathrm{orb}^\mathrm{1D/3D}$ are the orbital separations in the 1D and 3D simulation. For the calculation of the MRD, only the plunge-in phase is considered, \ie, the phase during which the orbital separation changes dynamically\footnote{We do not simulate the initial phase where the binary star loses co-rotation, but rather start the simulation at the beginning of the plunge-in phase. This is achieved because the stars have already lost co-rotation according to our setup.}. The end of the plunge-in phase is determined by
\begin{equation}\label{eq:end_plunge_in}
    \left\langle -\frac{\dot{a}_\mathrm{orb}}{a_\mathrm{orb}} P_\mathrm{orb} \right\rangle_\mathrm{orbit} < 0.01,
\end{equation}
where the time average is computed over one full orbit. This defines the end of the plunge-in phase when the average change in orbital separation during one orbit is less than \qty{1}{\%}. At the beginning of the CE phase in the 3D simulations, a transition of the initially circular orbit towards the dynamical spiral-in is observed. These first \qty{100}{d} are omitted in the MRD calculations in \eqtnref{eq:MRD}.

The best fit of the 1D simulation to the 3D simulation is determined manually. We find the best-fitting simulation by computing 1D CE models with varying values for \Cd and \Ch. For each simulation, we find $\Delta t_\mathrm{min}$ and then compare by eye the spiral-in curve and the mass fraction of ejected envelope to the 3D simulation.

\vb{We tested the above described method at different spacial and temporal resolutions and present the results in Appendix~\ref{apdx:resolution_study}.}

%
%
\section{Results for mass ratio $q=0.25$}\label{sec:results_q0.25}

In this section, we show the results of our 1D CE simulations with mass ratio $q=0.25$ and compare our results with the 3D hydrodynamic simulations of \citet{Sand2020}. First, we present the best-fitting 1D CE simulations in \sectref{sec:res_best-fit}. We then show how the drag-force parameter \Cd and the heating parameter \Ch affect the outcome of the simulations in \sectref{sec:res_varying_Cd_Ch}.  In \sectref{sec:res_dyn_evolution}, the dynamical processes taking place in the envelope are discussed, before we describe recombination processes in \sectref{sec:res_recombination}. The energy budget during the CE phase and the evolution of the drag force are shown in Sects.~\ref{sec:res_energy} and \ref{sec:res_Fdrag} respectively.

\subsection{Best fitting 1D simulation for $q=0.25$}\label{sec:res_best-fit}
\begin{figure}
    \resizebox{\hsize}{!}{\includegraphics{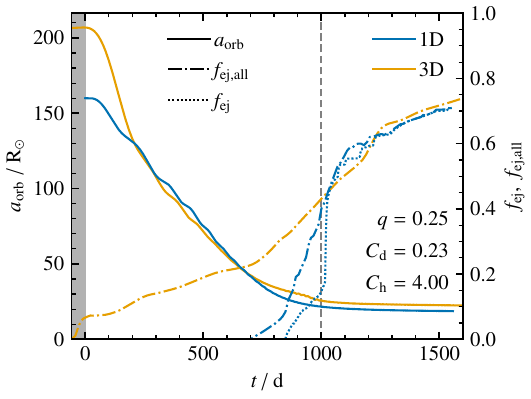}}
    \caption{Best fitting 1D CE simulation for $q=0.25$ when comparing to the 3D CE simulation of \citet{Sand2020}. Full lines show the orbital separation, and the \vb{dotted} and \vb{dash-dotted} lines show the mass fraction of ejected envelope $f_\mathrm{ej}$ and $f_\mathrm{ej,all}$ respectively. The results of the 3D simulations are shown in orange, and the results from this work are shown in blue. The vertical gray-dashed line indicates the end of the dynamical plunge-in phase as determined by \eqtnref{eq:end_plunge_in}. All the results of the 3D simulation are shifted by $\Delta t_\mathrm{min} = \qty{- 58}{d} $ to compensate for the difference in initial separation. The gray shaded region shows $t < \qty{0}{d}$ for which we cannot compare the 1D simulation to the 3D simulation.}
    \label{fig:best_fit_q025}
\end{figure}

The spiral-in curve of the best-fitting 1D simulation with parameters $\Cd = 0.23$ and $\Ch = 4.0$ for a mass ratio of $q = 0.25$ is shown in \figref{fig:best_fit_q025}. The spiral-in curve of the 3D benchmark model of \citet{Sand2020} for the same mass ratio is also shown. We find $\Delta t_\mathrm{min}= - \qty{58}{d} $ to best account for the difference in the initial separation (see~\sectref{sec:compare_3D}). The plunge-in phase is well reproduced by the 1D simulation. The orbital separation of the 1D simulation after the plunge-in phase is smaller than in the 3D simulation. In both the 1D and the 3D simulations, the post-plunge-in orbit shows a non-zero eccentricity. The eccentricity $e$ of the orbit can be approximated by
\begin{equation}\label{eq:eccentricity}
    e = \frac{A - P}{A + P}
\end{equation}
where $A$ is the apastron distance and $P$ is the periastron distance. This approximation is valid as long as the orbital separation is not changing significantly during one orbit, \eg, during the post-plunge-in phase of the CE evolution. In the 1D simulation, we find an eccentricity of 0.005, which is approximately equal to the eccentricities of 0.006 in the 3D simulation (for a summary, see \tabref{tab:summary}). In both cases, the eccentricity is evaluated at the end of the plunge-in phase as determined by \eqtnref{eq:end_plunge_in}, and averaged over 5 orbits. 

The fraction of unbound envelope material $f_\mathrm{ej}$ is also shown in \figref{fig:best_fit_q025}. For the 1D simulation, two descriptions ($f_\mathrm{ej}$ and $f_\mathrm{ej,all}$) are used to determine the fraction of the unbound envelope (\sectref{sec:compare_3D}). In the 3D simulation, the envelope ejection starts at $t < \qty{0}{d}$ and reaches 0.1 at $t = \qty{0}{d}$ (\figref{fig:best_fit_q025}). As the companion starts to enter the envelope, it causes one large spiral arm where mass is immediately ejected \citep{Sand2020}. In the 1D simulation, envelope ejection sets in later during the simulations. When considering $f_\mathrm{ej,all}$, the envelope ejection starts at $t = \qty{700}{d}$ compared to $t = \qty{850}{d}$ based on $f_\mathrm{ej}$. This is expected since $f_\mathrm{ej,all}$ captures all the unbound mass that is considered for $f_\mathrm{ej}$ and, additionally, all layers with $v_r > v_\mathrm{esc}$ inside the envelope. At the end of the 1D simulation, about $70\,\%$ of the envelope is unbound, as well as $f_\mathrm{ej} \approx f_\mathrm{ej,all}$, \ie, there are only unbound layers at the outer boundary of the CE but not within the CE. It is important to note that the slopes of the fraction of the ejected envelope, \ie, the envelope-ejection rate, at the end of the simulations are similar in both the 1D and 3D case, possibly suggesting a similar ejection mechanism. The envelope ejection in the 1D simulations relies on the energy released from the recombination of hydrogen and helium and will be analyzed in more detail in \sectref{sec:res_dyn_evolution}.

The 1D simulation ended because of too short time-steps. However, it seems appropriate to assume that the envelope ejection will be sustained if the simulation is run for longer, possibly leading to a full envelope ejection. At the end of the 3D simulation at \qty{4000}{d}, \qty{91}{\%} of the envelope is ejected (cf.\ Table 4 in \citealp{Sand2020}).

Although there are small quantitative differences between the 1D CE simulation in this work and the 3D CE simulations of \citet{Sand2020}, it is remarkable that it is possible to achieve such similar results with our 1D model.

\subsection{Role of \Cd and \Ch for the orbital spiral-in and envelope ejection}\label{sec:res_varying_Cd_Ch}
\begin{figure*}
    \includegraphics[width=17cm]{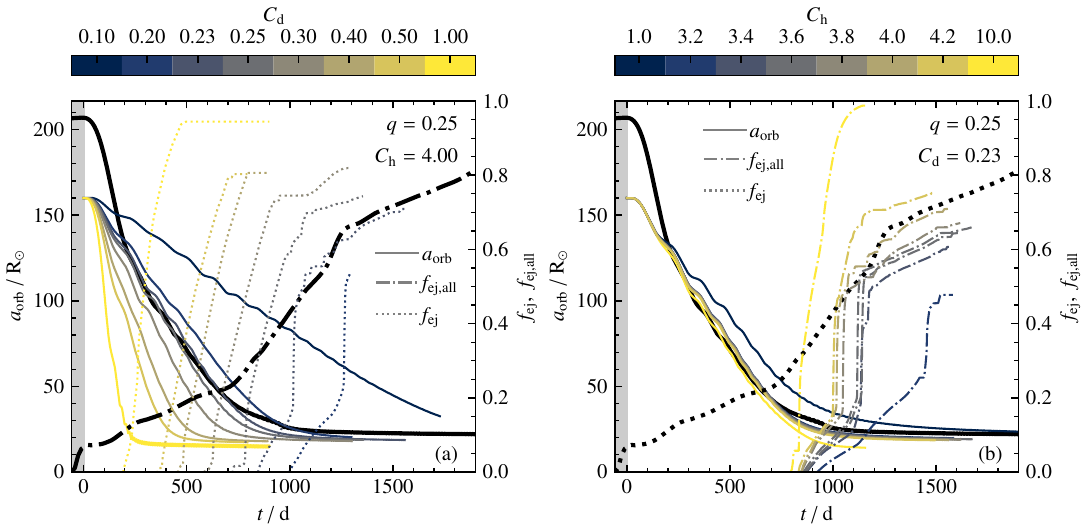}
    \caption{Effects of \Cd and \Ch on the spiral-in curves $a_\mathrm{orb}$ and the envelope-ejection fraction $f_\mathrm{ej}$. In panel (a), the heating parameter is held constant at $\Ch = 4.00$ while varying the drag-force parameter \Cd. In panel (b), the drag-force parameter is held constant at $\Cd = 0.23$ while varying the heating parameter \Ch. The thick black lines in both panels show the results from the 3D hydrodynamic CE simulation of \citet{Sand2020}, which are shifted by $\Delta t_\mathrm{min} = - \qty{58}{d}$, \ie, the time shift determined using the best-fitting 1D simulation with $\Cd = 0.23$ and $\Ch = 4.00$.}
    \label{fig:compare_Cd_Ch}
\end{figure*}

In \figref{fig:compare_Cd_Ch}, we show how \Cd and \Ch affect the orbital separation and the mass fraction of the ejected envelope by varying \Cd between $0.1$ and $1.0$ at the best-fit, constant $\Ch = 4.00$ (\figref{fig:compare_Cd_Ch}a), and varying \Ch between $1.0$ and $10.0$ at the best-fit, constant $\Cd=0.23$ (\figref{fig:compare_Cd_Ch}b). For simplicity, we only show $f_\mathrm{ej}$ in \figref{fig:compare_Cd_Ch}. However, we find for all simulations that $f_\mathrm{ej} \approx f_\mathrm{ej,all}$ during the late phases of the simulations, where the envelope-ejection rate is constant (Figs.~\ref{fig:best_fit_q025} and \ref{fig:higher_q}).

The slope of the spiral-in is mostly determined by the strength of the drag force via \Cd (\figref{fig:compare_Cd_Ch}a). When varying \Cd between $0.1$ and $1.0$ at fixed \Ch, the spiral-in timescale decreases from more than \qty{1500}{d} to less than \qty{200}{d}. Additionally, the envelope ejection starts earlier for larger \Cd while also ejecting a larger fraction of the envelope. The start of the rapid envelope ejection occurs at similar orbital separations of about $\qty{30}{\rsun}$ throughout the different simulations. The post-plunge-in separation changes only slightly when varying the drag-force parameter \Cd.  

We find that the heating parameter \Ch plays an important role in determining the mass fraction of the ejected envelope, but not so much in setting the post-plunge-in separation (\figref{fig:compare_Cd_Ch}b). Increasing the heating parameter causes a slightly deeper spiral-in of the companion and a higher fraction of ejected envelope. The behavior of $f_\mathrm{ej}$ for $\Cd = 0.23$ and varying \Ch between 3.2 and 4.2 is noteworthy. In this range, a changing \Ch results in a large change in $f_\mathrm{ej}$ while the orbital separation at the end of the plunge-in changes only between \qty{22.9}{\rsun} for $\Ch = 3.2$ and \qty{21.3}{\rsun} for $\Ch = 4.2$. For the same models, the envelope-ejection curves show a rapid ejection event at the beginning, after which a more steady envelope ejection settles in. In this second phase, the envelope-ejection rates seem to be the same for all the simulations, which are also in agreement with the 3D simulation. This is another indicator, that the envelope ejection, especially in the later phase of the simulation, is mainly caused by recombination processes because the recombination rate is comparable between the models (\sectref{sec:res_recombination}). Only the timing and strength of the initial rapid ejection event is changing when varying \Ch. Therefore, the heating parameter \Ch is important to set the envelope ejection in the 1D CE simulation.

From these experiments, we conclude that the drag-force parameter \Cd mostly affects the orbital separation during the plunge-in phase by determining the spiral-in timescale. The final mass-fraction of the ejected envelope is determined by both the heating parameter \Ch and the drag-force parameter \Cd. The orbital separation after the plunge-in phase changes only slightly when varying \Cd and \Ch. This suggests, that there is a more fundamental principle that sets the post-plunge-in separation and that does not depend on the details of the plunge-in-phase. When comparing the 1D model to the 3D simulation, we cannot match the results with only one of the two parameters. Therefore, 2 parameters are needed in 1D CE models similar to ours to reproduce the spiral-in curve as well as the envelope ejection.

\subsection{Dynamical evolution of the envelope}\label{sec:res_dyn_evolution}
\begin{figure*}
    \includegraphics[width=17cm]{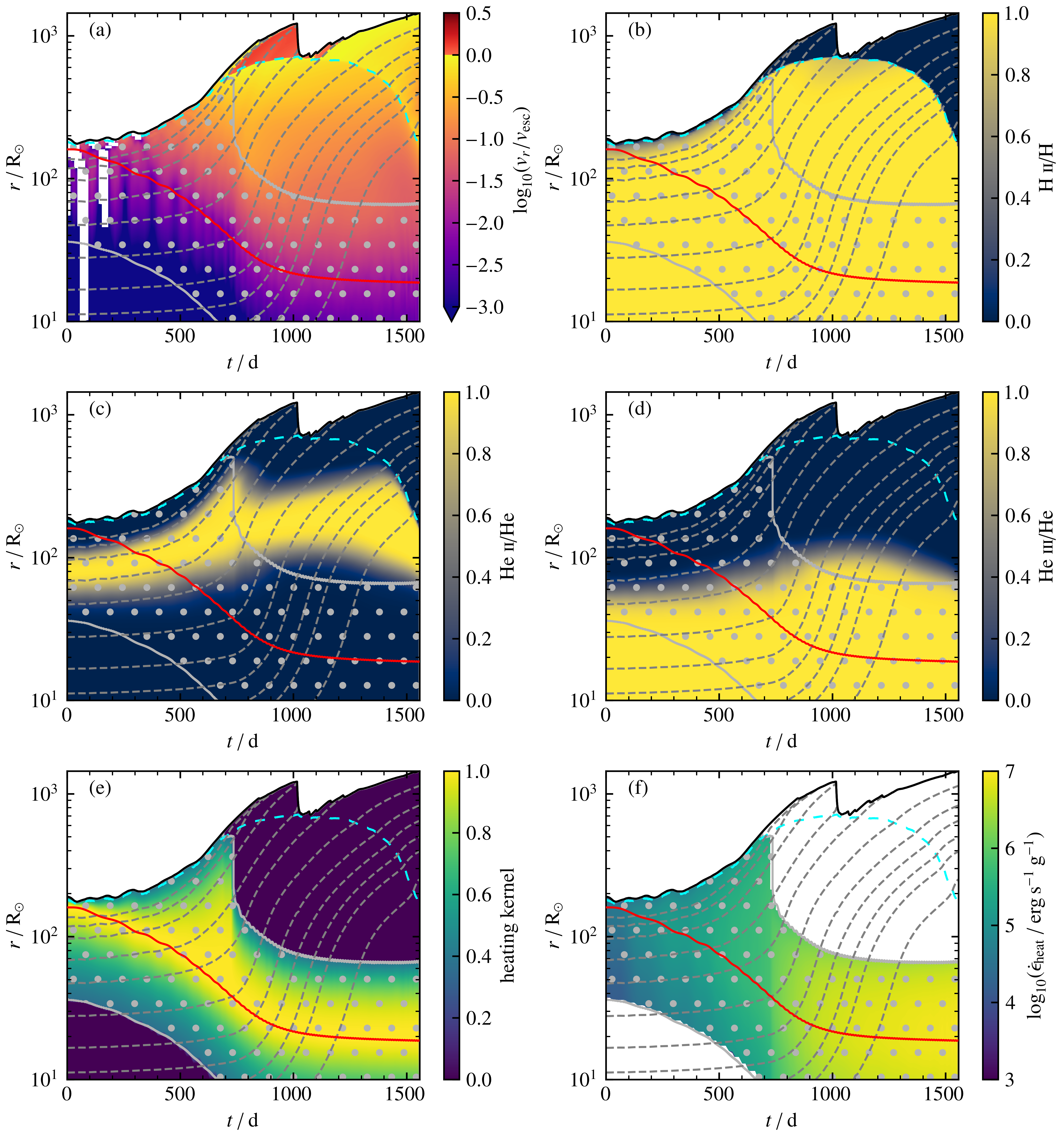}
    \caption{Kippenhahn diagrams of the simulation with $q = 0.25$, $\Cd = 0.23$ and $\Ch = 4.0$ showing the radial velocity in terms of the local escape velocity (panel a), where layers colored in red have a radial velocity larger than the local escape velocity, the ionization fractions of \ion{H}{ii}, \ion{He}{ii} and \ion{He}{iii} (panels b, c and d respectively), the heating kernel as described in \eqtnref{eq:edot_heat} (panel e), and the specific heating rate (panel f). The red line indicates the orbital separation between the companion and the core of the giant star. The gray-dotted region is heated during the CE simulation. The cyan dashed line shows \Rtau which traces the \recrad. The gray dashed lines represent envelope-mass-fractions of 0.95, 0.9, 0.8, 0.6, 0.4, 0.2, 0.1, 0.06, 0.04, 0.03 from the surface toward the center, and visualize the expansion of the envelope. The white patches in panel (a) show layers that have an inward (negative) radial velocity.}
    \label{fig:kipp1}
\end{figure*}

\begin{figure*}
    \includegraphics[width=17cm]{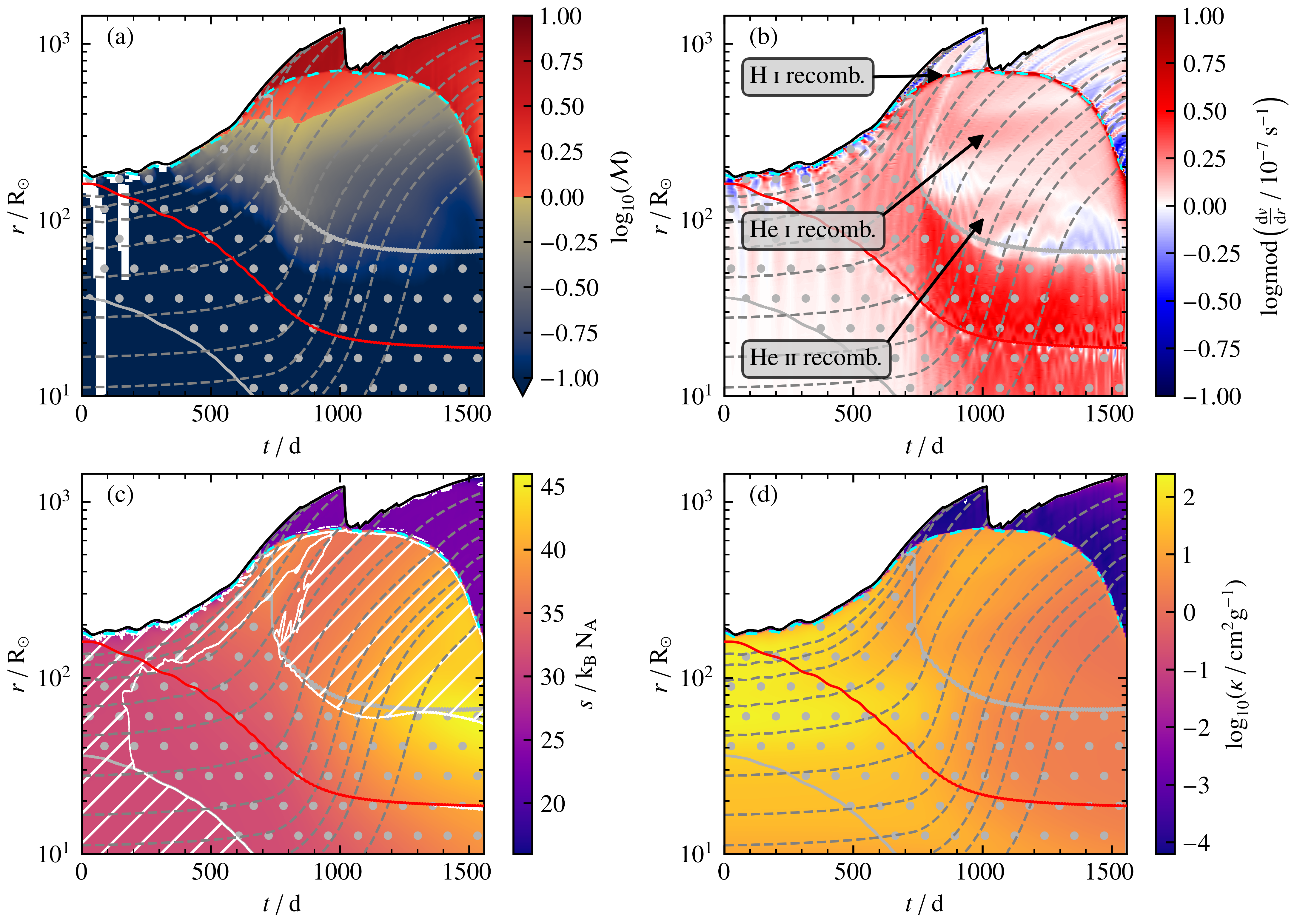}
    \caption{Kippenhahn diagrams similar to \figref{fig:kipp1} of the simulation with $q = 0.25$, $\Cd = 0.23$ and $\Ch = 4.0$ showing the Mach number $\mathcal{M} = v_r/c_\mathrm{s}$, velocity divergence $\mathrm{d}v / \mathrm{d}r$, specific entropy $s$, and opacity $\kappa$. The white hatched zones in panel (c) indicate convective mixing. For better visualizing the velocity divergence, the log modulus transformation, $\mathrm{logmod}(x) = \mathrm{sgn}(x) \log_{10}(|x| + 1)$, is used \citep{John1980}. The arrows in panel (b) roughly indicate the recombination zones of $\ion{H}{i}$, $\ion{He}{i}$ and $\ion{He}{ii}$ (see also panels (b), (c) and (d) in \figref{fig:kipp1}).}
    \label{fig:kipp2}
\end{figure*}

The dynamical evolution of the envelope is shown by several Kippenhahn diagrams in Figs.~\ref{fig:kipp1}~and~\ref{fig:kipp2}. During the first ${\sim}\, \qty{200}{d}$, the envelope oscillates radially. These oscillations are artifacts of the change to the hydrodynamic mode in \mesa as well as the change in the boundary condition, which are not yet fully damped during the relaxation run (see~\sectref{sec:relaxation}). As the velocities in these oscillations are small compared to the expected expansion velocity which is comparable to the escape velocity, they do not affect the further evolution of the CE.

After this initial phase, the envelope begins to expand, as more and more heat is injected into the outer layers (see~\figref{fig:energy_budget}). The partial ionization zones of hydrogen and helium are expanding together with the envelope, as can be seen in Figs.~\ref{fig:kipp1}b, c, and d, and no recombination is happening yet.

For $t < \qty{700}{d}$, we find $r_\mathrm{max}^\mathrm{heat} = R_1$, \ie, the upper boundary of the heating zone coincides with the outer boundary of the simulation domain. Once $r_\mathrm{max}^\mathrm{heat} < R_1$ for $t > \qty{700}{d}$, the specific heating rate $\dot{\epsilon}_\mathrm{heat}$ increases by about one order of magnitude because the mass of the material that is heated is decreasing significantly (Figs.~\ref{fig:kipp1}e and f). As the upper heating radius moves deeper in the envelope, the partial ionization zones of singly ionized hydrogen and helium are no longer heated and recombination sets in. The beginning of recombination and the end of heating in these zones are causally connected, as the heating provides an energy source that keeps the atoms ionized. This is also the reason why \Ch determines the ejection of the envelope so strongly.

The layer in which hydrogen recombination takes place moves outwards in radius at a typical optical depth of $\tau = 10$ (\Rtau), hereafter referred to as the \recrad, below the outer boundary of the simulation. We find that \Rtau stays always close to the hydrogen recombination front because, above the hydrogen recombination layer, the opacity drops dramatically (\figref{fig:kipp2}d). In fact, this layer expands faster than the local escape velocity (\figref{fig:kipp1}a). This demonstrates that the energy released from hydrogen recombination provides an important acceleration mechanism in expanding and ejecting the envelope. Even though the recombination spatially takes place close to the photosphere (with $\tau\sim 1$), it is still in the optically thick region where the recombination photons cannot freely escape\footnote{At an optical depth of $\tau=10$, only $e^{-10} \approx \qty{0.005}{\%}$ of the photon are expected to \vb{escape} without any scattering event, compared to $\,\approx\qty{37}{\%}$ at the photosphere around $\tau \sim 1$. Initially, $R_{\tau=1}$ and $R_{\tau=10}$ are spatially separated by $\lesssim \qty{10}{\rsun}$ which increases to $\gtrsim \qty{1000}{\rsun}$ around $t=\qty{1500}{d}$. This illustrates the significant difference between $\tau=1$ and $\tau=10$. It is worth noting that in 3D simulations, this transition region is difficult to resolve.} (also see \citealt{Ivanova2016,Clayton2017}).
On top of this fast layer at $\qty{700}{d} \leq t \leq \qty{1000}{d}$, there is a slower-moving layer that prevents the fast-expanding layer from becoming unbound and being removed. As the fast material crashes in the slower layer on top, a shock wave is produced, seen by the large negative velocity divergence $\mathrm{d}v / \mathrm{d}r$ in \figref{fig:kipp2}b.

Layers exceeding $v_\mathrm{esc}$ reach the outer simulation boundary at $t = \qty{1000}{d}$. Then, a rapid mass-loss event removes ${\sim}\, \qty{30}{\%}$ of the envelope mass in a short period of time, followed by continuous steady envelope ejection. The sudden mass-loss event can also be seen in \figref{fig:best_fit_q025}, as the mass fraction of ejected envelope $f_\mathrm{ej}$ increases rapidly at $t = \qty{1000}{d}$.

We find that \Rtau stays constant for $700 - \qty{1200}{d}$, after which \Rtau decreases as the hydrogen recombination front moves to smaller radii. At about $t = \qty{1400}{d}$, the singly ionized helium-recombination front is located at the \recrad, which means that hydrogen and singly ionized helium recombine at approximately the same physical location. The released recombination energy accelerates the envelope material, which can be seen by the increase in the slope of the lines of constant mass in \figref{fig:kipp1}.

Along the \recrad, there is a layer of positive velocity divergence $\mathrm{d}v / \mathrm{d}r$ (\figref{fig:kipp2}b). This is another indicator that the energy from hydrogen recombination contributes to accelerate the envelope. The recombination of singly and doubly-ionized helium also causes layers of positive velocity divergence but with a smaller magnitude compared to the recombination of hydrogen. Because the partial ionization zones of singly and doubly-ionized helium are more radially extended than the partial ionization zone of hydrogen, the increase in the velocity divergence is less pronounced. Once the upper heating radius decreases at $t = \qty{700}{d}$, the velocity divergence inside the heating zone, \ie, all layers with $r_\mathrm{min}^\mathrm{heat} \leq r \leq r_\mathrm{max}^\mathrm{heat}$, increases significantly (\figref{fig:kipp2}b). This is an immediate consequence of the increase in the specific heating rate as a result of the decrease in the total envelope mass that is heated (\figref{fig:kipp1}f). Therefore, the localized heating source causes the envelope to expand rapidly. At the same time, there is an outward-traveling feature defined by a lower velocity divergence compared to the surroundings (\figref{fig:kipp2}b). This feature is launched at $t = \qty{750}{days}$ at the top of the heating zone and reaches the outer boundary at $t = \qty{1000}{d}$. The cause of this feature, possibly a pressure/acoustic wave, is likely connected to the decrease in the size of the heating layers. 

The neutral layers outside the \recrad expand supersonically (\figref{fig:kipp2}a). The sound speed of the outer layers is significantly lower due to adiabatic and photon cooling in the neutral layers (\figref{fig:kipp2}c). These outward-moving layers show alternating positive and negative velocity divergence (\figref{fig:kipp2}b), \ie, there are alternating layers with faster and slower expansion velocities compared to the average expansion velocity of the envelope. The details of the origin of this pattern are unclear. The large negative velocity divergence at the end of the simulation between $ 200 - \qty{300}{\rsun}$ indicates a shock wave, but we cannot observe any direct consequences arising from the shock.

The white hatching in \figref{fig:kipp2}c shows layers in the CE that are unstable against convection. Initially, the entire envelope is convective, \ie, the entropy gradient is less than or equal to zero. Heating stops the convective energy transport after about \qty{200}{d}, at the same time as the envelope begins to expand. For $t < \qty{200}{d}$, the injected energy is transported almost instantaneously to the outer boundary, without affecting the entropy structure of the envelope. As the heating rate increases with time (\figref{fig:kipp1}f) because of the increase in the drag force, the injected energy stops convection. Consequently, the energy cannot be transported away by convection but rather leads to an expansion of the envelope. As energy is injected into the heating zone, the specific entropy in the heating zone increases. The gradient of the heating rate $\dot{\epsilon}_\mathrm{heat}$ is positive between $r_\mathrm{min}^\mathrm{heat}$ and $a_\mathrm{orb}$ (Eq.~\ref{eq:edot_heat}). Hence, the entropy gradient is also expected to be positive for $r_\mathrm{min}^\mathrm{heat} \leq r \leq a_\mathrm{orb}$, implying that these layers become stable against convection. While for $a_\mathrm{orb} \leq r \leq r_\mathrm{max}^\mathrm{heat}$ the heating rate decreases (Eq.~\ref{eq:edot_heat}) and most of these layers are also stable against convection for $t > \qty{750}{d}$. They have a positive entropy gradient because the heat/entropy of the inner layers is transported outwards. Outside the \recrad, the transport of energy by radiation is more efficient than convection due to the lower opacity. Some smaller convective zones are visible close to the companion and the outer boundary of the CE at the end stages of the simulations. They are not expected to affect the evolution of the CE because of their limited radial extent.

At the end of the simulation, the opacity close to the outer boundary of the CE increases (\figref{fig:kipp2}f). This increase is caused by the appearance of hydrogen molecules that can form at such low temperatures and densities. At these temperatures, dust formation close to the outer boundary might be possible as well, but it is not included in our simulations (dust formation is expected at a temperature ${\sim}\, 1200 - \qty{2100}{K}$; \citealt{Iaconi2020}).

\subsection{Recombination energy}\label{sec:res_recombination}
\begin{figure}
    \resizebox{\hsize}{!}{\includegraphics{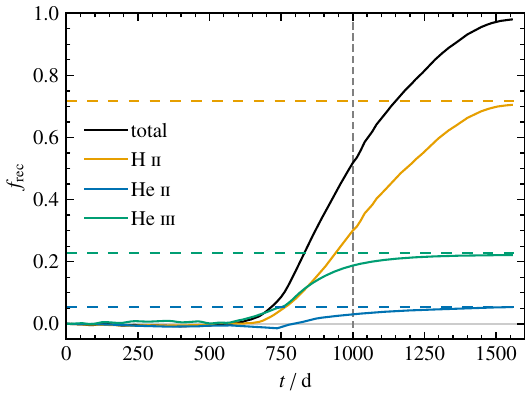}}
    \caption{Fraction of released recombination energy $f_\mathrm{rec}$ and contributions of hydrogen (\ion{H}{ii}), singly ionized helium (\ion{He}{ii}) and doubly ionized helium (\ion{He}{iii}) for the simulation with $q = 0.25$, $\Cd = 0.23$ and $\Ch = 4.0$. The total fraction of released recombination energy is shown in black. The horizontal colored-dashed lines show the initially available recombination energy in each species. The vertical gray-dashed line indicates the end of the dynamical plunge-in phase as determined by \eqtnref{eq:end_plunge_in}.}
    \label{fig:recombination}
\end{figure}
To study the effects of recombination on envelope ejection, we calculate the amount of released recombination energy. For this, the ionization potentials of hydrogen (\ion{H}{ii}), singly-ionized helium (\ion{He}{ii}) and doubly-ionized helium (\ion{He}{iii}) are taken from \citet{NIST_ASD},
\begin{align}
    E_\ion{H}{ii} & \equiv E_{\ion{H}{i} \rightarrow \ion{H}{ii}} = 13.5984 \, \mathrm{eV} \\
    E_{\ion{He}{ii}} & \equiv E_{\ion{He}{i}\rightarrow \ion{He}{ii}} = 24.5874 \, \mathrm{eV} \\
    E_{\ion{He}{iii}} & \equiv E_{\ion{He}{ii}\rightarrow \ion{He}{iii}} = 54.4178 \, \mathrm{eV} 
\end{align}
and the atomic masses from \citet{Prohaska2022},
\begin{align}
    m_\mathrm{H} &= 1.0080 \, \mathrm{amu} \\
    m_\mathrm{He} &= 4.0026 \, \mathrm{amu}.
\end{align}
The potentially available energy from the recombination of hydrogen and helium, usually referred to as the recombination energy, is then given by 
\begin{multline}\label{eq:Erec_int}
    E_\mathrm{rec} = \int\limits_{M_\mathrm{1,core}}^{M_1}\left[ \frac{X}{m_\mathrm{H}}E_{\ion{H}{ii}} f_{\ion{H}{ii}} \right .\\
    \left. + \frac{Y}{m_\mathrm{He}}\Big(E_{\ion{He}{ii}} f_{\ion{He}{ii}} + (E_{\ion{He}{iii}} + E_{\ion{He}{ii}}) f_{\ion{He}{iii}} \Big) \right]\mathrm{d}m,
\end{multline}
where $X$ and $Y$ are the mass fractions of hydrogen and helium, and $f_{\ion{H}{ii}}$, $f_{\ion{He}{ii}}$ and $f_{\ion{He}{iii}}$ are the ionization fractions of \ion{H}{ii}, \ion{He}{ii} and \ion{He}{iii} respectively\footnote{The envelope material that is removed during the simulation has fully recombined before reaching velocities larger than the escape velocity (\figref{fig:kipp1}). Thus, the envelope ejection does not remove any potential recombination energy from the system.}. The recombination energies of elements more massive than helium are not taken into account, because their contribution to the total recombination energy is negligible \citep{Ivanova2016}.

In \figref{fig:recombination}, we show the fractions $f_\mathrm{rec,total}$ and $f_{\mathrm{rec},i}$ of released recombination energy
\begin{align}
    f_\mathrm{rec,total} &= \frac{E_\mathrm{rec, ini} - E_\mathrm{rec}}{E_\mathrm{rec, ini}},  \\
    f_{\mathrm{rec},i} &= \frac{E_{\mathrm{rec, ini}\vbmath{, i}} - E_{\mathrm{rec},i}}{E_\mathrm{rec, ini}} \; \mathrm{with} \; i={\ion{H}{ii},\, \ion{He}{ii},\, \ion{He}{iii}},
\end{align}
where $E_\mathrm{rec,i}$ is computed by integrating each term in \eqtnref{eq:Erec_int} individually.

The released recombination energy is negligible for $t < \qty{700}{d}$. This can be understood with the help of \figref{fig:kipp2}, which shows that the partial ionization zones of hydrogen and helium remain at constant mass coordinates during this time. At the end of the simulation, more than $98\,\%$ of the initially available recombination energy is released, and about $70\,\%$ of the released energy is due to hydrogen recombination. The reason for this is the high abundance of hydrogen in the envelope ($X_\mathrm{ini} = 0.7$). For $\qty{600}{d }\leq t \leq \qty{750}{d}$, we find $f_{\mathrm{rec},\ion{He}{ii}} < 0$, because doubly-ionized helium recombines to form singly-ionized helium such that the recombination energy available from singly-ionized helium is larger than the initial recombination energy stored in singly-ionized helium.

Our calculations show that recombination energy is likely to provide an important contribution to the envelope ejection process. Hydrogen recombination occurs well below the photosphere (at a typical optical depth $\tau=10$), where photons cannot escape directly. This results in a positive velocity divergence at the location where hydrogen recombines, causing further acceleration of the envelope (\figref{fig:kipp2}b).
The recombination front of singly-ionized helium is deeper in the envelope compared to the recombination front of hydrogen. Around this recombination zone, the velocity divergence is positive. It is not as large as in the case of hydrogen recombination but much more extended radially. A similar behavior can be observed for the recombination zone of doubly-ionized helium. There is a layer of negative velocity divergence around the \ion{He}{iii} recombination zone for $t > \qty{1300}{d}$. As most of the doubly-ionized helium has already recombined at this point (see~\figref{fig:recombination}), the energy released might not be enough to further accelerate the envelope.

At the end of the plunge-in phase, about $50 \, \%$ of the available recombination energy is released, while only ${\sim}\, 40 \, \%$ of the envelope is ejected. Hereafter, the energy that drives the envelope ejection is mostly from recombination and no longer from the heating, as by definition the orbital separation decreases slowly in the post-plunge-in phase which means that the drag force and hence the heating are much lower than during the plunge-in phase (see also Sects.~\ref{sec:res_energy} and \ref{sec:res_Fdrag}). At the end of the simulation, more than $70 \, \%$ of the envelope is ejected, which suggests that the recombination energy contributes significantly to the ejection, especially in the post-plunge-in phase. For a comparison, at the end of the 3D simulation at $t=\qty{2500}{d}$, $\qty{91}{\%}$ of the envelope mass is ejected \citep{Sand2020}.

\subsection{Energy budget}\label{sec:res_energy}

\begin{figure}
    \resizebox{\hsize}{!}{\includegraphics{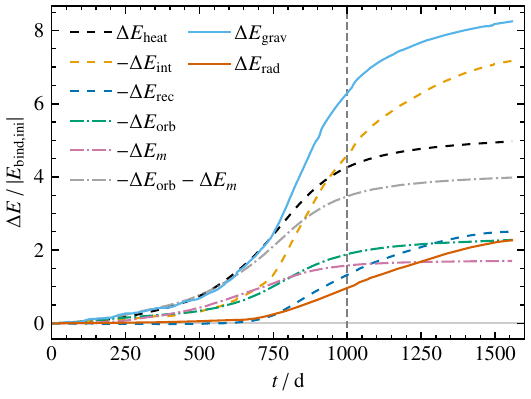}}
    \caption[]{\vb{Major energy} sources \vb{(dashed lines)} and sinks \vb{(full lines)} during the CE simulation with $q = 0.25$, $\Cd = 0.23$ and $\Ch = 4.0$. Shown are the total injected energy $\Delta E_\mathrm{heat}$, \vb{the change in internal energy $\Delta E_\mathrm{int}$, the total released recombination energy $\Delta E_\mathrm{rec}$ (the recombination energy is part of the internal energy and is only shown separately to highlight the contribution from recombination)}, the change in orbital energy $\Delta E_\mathrm{orb}$ according to \eqref{eq:E_orb}, the extra term $\Delta E_m$ defined by \citet{Yarza2022b}, \vb{the change in the gravitational energy $\Delta E_\mathrm{grav}$}, and the energy that is carried away by radiation $\Delta E_\mathrm{rad}$. The vertical gray-dashed line indicates the end of the dynamical plunge-in phase as determined by \eqtnref{eq:end_plunge_in}. The energies are shown in units of the initial binding energy of the envelope with $E_\mathrm{bind,ini} = \qty{-4.03e45}{erg}$\footnotemark. \vb{The lines showing contributions from $\Delta E_\mathrm{orb}$ and $\Delta E_\mathrm{m}$ are dash-dotted, as these show source terms for the heating $\Delta E_\mathrm{heat}$, but are not direct energy sources in the \mesa simulation.}}
    \label{fig:energy_budget}
\end{figure}

\vb{Major energy} sources and sinks during the 1D CE simulation are shown in \figref{fig:energy_budget}. The energies are shown in units of the initial binding energy of the envelope
\begin{equation}
    E_\mathrm{bind} = \int_{M_{1,\mathrm{core}}}^{M_1} \left( -\frac{G m}{r} + u \right) \, \mathrm{d}m ,
\end{equation}
where $u$ is the internal energy. The total energy injected via heat is determined from the drag force and the relative velocity by
\begin{equation}
    \Delta E_\mathrm{heat} = -\int \vec{F}_\mathrm{d} \cdot \vec{v}_\mathrm{rel} \, \mathrm{d}t.
\end{equation}
The source of the heating energy is the orbital energy of the two stars in the CE phase. The released orbital energy is classically given by 
\begin{equation}\label{eq:E_orb}
    \Delta E_\mathrm{orb} = -\frac{G M_{1,a_\mathrm{orb}} M_2}{2\, a_\mathrm{orb}} + \dfrac{G M_1 M_2}{2\, a_\mathrm{orb,ini}},
\end{equation}
where $M_{1,a_\mathrm{orb}}$ is the mass of $M_1$ enclosed by the orbit with separation $a_\mathrm{orb}$. From \figref{fig:energy_budget}, it is clear that $\Delta E_\mathrm{heat}$ is approximately a factor 2 larger than $ - \Delta E_\mathrm{orb}$. \vb{This discrepancy is expected in our CE formalism because we mix both 1D and 3D treatments of physical mechanisms. For example, we evolve the binary orbit in 3D but are bound to use 1D approximations to calculate the change in potential energy. Therefore, we cannot expect to conserve energy in our CE formalism, \ie, we do expect $\Delta E_\mathrm{heat} \neq -\Delta E_\mathrm{orb}$.} 

\vbre{In addition, \eqtnref{eq:E_orb} only applies to systems, where the envelope is completely ejected, and the ejecta has zero velocity at infinity. In our simulations, the above-mentioned criteria are not fulfilled, and, therefore, \eqtnref{eq:E_orb} is insufficient to capture all the relevant effects.}
\vb{This means, that there is a need for extra terms \vbre{in \eqtnref{eq:E_orb} to correctly calculate $\Delta E_\mathrm{orb}$} in 1D CE treatments similar to ours.} \vbre{One such term arises from the change in potential energy because of the change in the enclosed mass. In the limit $M_2 \ll  M_{1,a_\mathrm{orb}}$, \citet{Yarza2022b} suggest this term to be}
\begin{equation}
    \Delta E_m = G M_2 \int_{a_\mathrm{orb,ini}}^{a_\mathrm{orb}} \frac{1}{r}\frac{\mathrm{d}M_{1,r}}{\mathrm{d}r}\, \mathrm{d}r,
\end{equation}
where $M_{1,r}$ is the mass of the giant star enclosed by the radius $r$. \vb{We test \vbre{the effect of $\Delta E_m$ to \eqtnref{eq:E_orb}} on our simulation, to see if it can resolve the observed mismatch between $\Delta E_\mathrm{heat}$ and $- \Delta E_\mathrm{orb}$.} Even when considering \vbre{the contribution of $\Delta E_m$ to} the \vbre{change in} orbital energy, there is more energy injected in the envelope than is released (\figref{fig:energy_budget}). \vbre{We note that in our simulation the condition $M_2 \ll  M_{1,a_\mathrm{orb}}$ is not fulfilled. Further reasons for the discrepancy are} \vb{that the change in the orbital energy is not only caused by the change in the orbital separation but also by the envelope expansion. As the envelope expands, the mass enclosed by the orbit $M_{1,a_\mathrm{orb}}$ decreases and thus the potential energy of the companion. This means that the orbital energy of a companion at a constant $a_\mathrm{orb}$ inside an expanding envelope increases because of the decrease in the enclosed mass by the orbit. This effect is not included in the term $\vbmathre{\Delta E_m}$ proposed by \citet{Yarza2022b}.}

\footnotetext{The definition of the core mass $M_\mathrm{1,core}$ is arbitrary and can affect the value of the binding energy hugely \vb{\citep{Tauris2001}}. For consistency, we use the definition of $M_\mathrm{1,core} = \qty{0.545}{\msun}$ of \citet{Sand2020} throughout this paper. When using $M_\mathrm{1,core} = \qty{0.540}{\msun}$, the helium core mass defined by a hydrogen mass-fraction less than 0.1, the binding energy of the envelope increases to $\qty{-5.67e46}{erg}$. Hence, the fractions $\Delta E / |E_\mathrm{bind,ini}|$ in \figref{fig:energy_budget} should not be used for absolute comparisons.}

Additionally, the energy loss via radiation at the outer boundary of the envelope is shown in \figref{fig:energy_budget} as well as the energy release from the recombination of hydrogen and helium. The losses from radiation are about half of the energy injected via heat. Both the losses via radiation and the energy gain from recombination start to become significant for $t > \qty{700}{d}$, after which they increase almost synchronously. This does not mean that the energy released from recombination is immediately radiated away, but rather contributes to accelerate the envelope material, \eg, seen by the kink in the lines of constant mass at the hydrogen recombination front (\figref{fig:kipp1}b) as well as by the positive velocity divergence (\figref{fig:kipp2}b). This acceleration can even cause the layers to expand faster than the local escape velocity (\figref{fig:kipp1}a).

\subsection{Drag force evolution}\label{sec:res_Fdrag}
\begin{figure}
    \resizebox{\hsize}{!}{\includegraphics{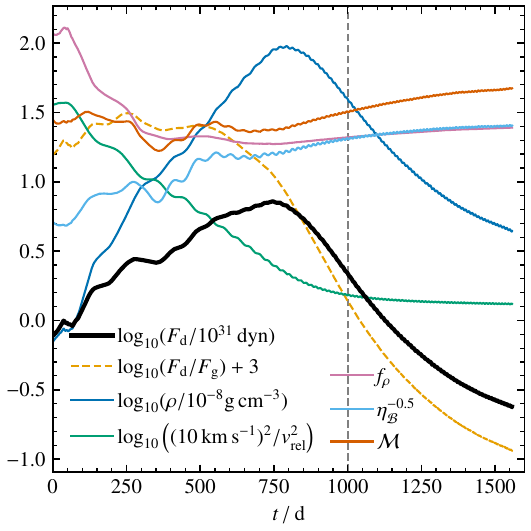}}
    \caption{Time evolution of the drag force and its components for $q=0.25$, $\Cd = 0.23$ and $\Ch=4.0$. The visualization of these quantities represents how each of them enters the drag force according to \eqtnref{eq:F_d}. During the whole simulation, the drag force is in the non-linear regime, \ie, $\mathcal{M} > 1.01$ and $\eta_\mathcal{B} > 0.1$. Additionally, the ratio of the drag force and gravitational force $F_\mathrm{g}$ is shown. The vertical gray-dashed line indicates the end of the dynamical plunge-in phase as determined by \eqtnref{eq:end_plunge_in}.}
    \label{fig:drag_force}
\end{figure}

The drag force acting on the companion is calculated by \eqtnref{eq:F_d} following the results of \citet{Kim2010} and \citet{Kim2007}. In \figref{fig:drag_force}, we show the individual components that enter the drag force as well as the ratio of the drag force and the gravitational force $F_\mathrm{g}$. The drag force increases for $t < \qty{750}{d}$. From \figref{fig:drag_force} it is apparent, that the drag force is mostly influenced by the density $\rho$ and the relative velocity $v_\mathrm{rel}$ as they vary the most compared to the other components of the drag force. For $t < \qty{750}{d}$, the density increases which causes the drag force to increase. The relative velocity increases throughout the entire simulation because the orbital velocity increases with ${\sim}\, a_\mathrm{orb}^{-1/2}$ and the rotational velocity decreases with ${\sim}\, a_\mathrm{orb}$ as the orbit separation $a_\mathrm{orb}$ decreases (compare \eqtnref{eq:v_rel}). Hence, the drag force, which is proportional to $v_\mathrm{rel}^{-2}$, decreases. For $t < \qty{750}{d}$, the increase of $\rho$ dominates over the decrease of $v_\mathrm{rel}^{-2}$, causing the drag force to increase. For $t > \qty{750}{d}$, the drag force decreases, because both the $\rho$ and $v_\mathrm{rel}^{-2}$ decrease. 

The ratio $F_\mathrm{d} / F_\mathrm{g}$ is an indicator for the relative strength of the drag force compared to the gravitational force, \eg, a larger value of $F_\mathrm{d} / F_\mathrm{g}$ shows a high drag force and hence we expect a larger orbital decay compared to a lower value of $F_\mathrm{d} / F_\mathrm{g}$. At $t = \qty{250}{d}$, $F_\mathrm{d} / F_\mathrm{g}$ starts to decrease, indicating that the relative strength of the drag force \vb{decreases} (\figref{fig:drag_force}). This is at approximately at same time as the envelope starts to expand (\figref{fig:kipp1}). Initially, we find values for $F_\mathrm{d} / F_\mathrm{g}$ between $10^{-1.8}$ and $10^{-1.5}$. At $t = \qty{750}{d}$, the drag force peaks and then decreases. This change in behavior of $F_\mathrm{d}$ translates into the change in slope of $F_\mathrm{d} / F_\mathrm{g}$. At the end of the plunge-in phase at $t = \qty{1000}{d}$, we find $F_\mathrm{d} / F_\mathrm{g} = \qty{e-3}{}$, which decreases to \qty{e-4} at the end of the simulation. This shows, that indeed the drag force becomes small compared to gravity during the post-plunge-in phase, which is the main reason, why the dynamical plunge-in ends.

\section{Results for mass ratios $q=0.50$ and $q=0.75$}\label{sec:results_higher_q}

\begin{figure*}
    \includegraphics[width=17cm]{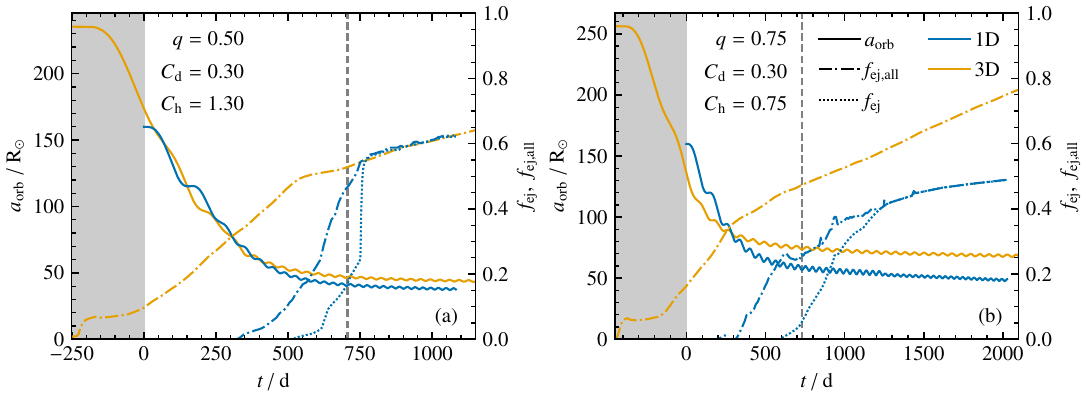}
    \caption{Similar to \figref{fig:best_fit_q025}, but showing the best fitting 1D simulation compared to the 3D simulation of \citet{Sand2020} for mass ratios $q=0.5$ in panel (a) and $q=0.75$ in panel (b).}
    \label{fig:higher_q}
\end{figure*}

The spiral-in and envelope-ejection curves for our 1D simulations with mass ratios $q=0.5$ and $q=0.75$ are shown in \figref{fig:higher_q}. For a mass ratio of $q = 0.5$, the spiral-in curve as well as the mass fraction of the ejected envelope capture well the results of the 3D simulation of \citet{Sand2020}. Similar to the case with mass ratio $q = 0.25$, the envelope-ejection rate in the later phases of the simulation is very similar to the 3D simulation. Additionally, both the 1D and the 3D simulation show a comparable post-plunge-in eccentricity of 0.020 and 0.017, respectively (see~\tabref{tab:summary}).

For a mass ratio of $q = 0.75$, however, it is more difficult to find values of \Cd and \Ch such that the 1D simulation reproduces well the spiral-in and the envelope ejection of the 3D simulation. Both the post-plunge-in orbital separation and the final envelope-ejection rate are lower than in the 3D simulation. This suggests that we have reached the limitations of our model as the companion for mass ratio $q = 0.75$ is no longer a small perturber, but rather comparable in mass to the initial AGB star and even more massive than its helium core of \qty{0.54}{\msun}. The eccentricity of the post-plunge-in orbit decreases after about $\vbmath{1250\, \mathrm{d}}$ (\figref{fig:higher_q}) because parts of the envelope material fall back below the orbit of the companion, causing a spike in the drag force that, in turn, alters the orbit.

\begin{table}
    \caption{Summary of the fitting parameters of the three 1D CE simulations. The orbital separations at the end of the plunge-in phase $a_\mathrm{orb,pl}$ are given for both the 1D and the 3D simulations. The eccentricities $e_\mathrm{pl}$ are determined for 1D and 3D simulations via \eqtnref{eq:eccentricity} at the end of the plunge-in phase. Both $a_\mathrm{orb,pl}$ and $e_\mathrm{pl}$ are averaged over 5 orbits.}
    \label{tab:summary}
    \centering
    \begin{tabular}{c c c c c c c}
        \hline \hline
        $q$ & \Cd & \Ch & $a_\mathrm{orb, pl}^\mathrm{1D}$ & $a_\mathrm{orb, pl}^\mathrm{3D}$ & $e_\mathrm{pl}^\mathrm{1D}$ & $e_\mathrm{pl}^\mathrm{3D}$ \\
         & & & $\left[\mathrm{R}_\odot\right]$ & $\left[\mathrm{R}_\odot\right]$ & & \\
        \hline
        0.25 & 0.23 & 4.00 & 21.2 & 25.2 & 0.005 & 0.006 \\
        0.50 & 0.30 & 1.30 & 40.1 & 47.6 & 0.020 & 0.017 \\
        0.75 & 0.30 & 0.75 & 57.2 & 76.7 & 0.034 & 0.018 \\
        \hline
    \end{tabular}
\end{table}

The best-fit parameters \Cd and \Ch as well as the post-plunge-in orbital separations and eccentricities of the three simulations with mass ratios $q = 0.25$, 0.5 and 0.75 are summarized in \tabref{tab:summary}. In all cases, the post-plunge-in orbital separation of the 1D simulations is smaller than that of the 3D simulation. Possible reasons are discussed in \sectref{sec:discussion}. For $q=0.25$ and 0.5, the eccentricity of both the 1D and the 3D simulations are comparable while for $q = 0.75$ the eccentricity in the 1D simulations is twice as large as in the 3D simulation. The drag-force parameter \Cd stays almost the same for all simulations with $\Cd {\sim}\, 0.25$. The heating parameter \Ch decreases with increasing mass ratio $q$. However, the results for a mass ratio $q = 0.75$ may not be accurate, as the simulation does not fit the 3D simulation as well as for the lower mass ratios. 

\vb{The results for $q=0.25$ discussed in \sectref{sec:results_q0.25} regarding the role of the recombination energy in ejecting the envelope, the behavior of the drag force and the energy budget qualitatively also apply for the simulations with higher mass ratios. We find that the recombination energy is important in driving the envelope ejection. The behavior of the drag force is mostly determined by the change in relative velocity and density. Both of which cause the drag force to drop at the end of the plunge-in phase.}

%
%
\section{Discussion}\label{sec:discussion}
In this section, we discuss our 1D CE method as well as the results that we described above. First, we show the limitations as well as the advantages of our 1D CE method in Sects.~\ref{sec:disc-limitations} and \ref{sec:dics-advantages} respectively. Then, we compare our 1D CE method to other proposed 1D methods for simulations of the CE phase within the context of 3D simulations (\sectref{sec:dics-comparison}). Finally, we physically motivate the necessity of the two free parameters used in our model (\sectref{sec:disc-two_parameters}).

\subsection{Limitations of the 1D CE model}\label{sec:disc-limitations}
In our 1D CE model, the drag force acts only on the companion and not on the core of the giant star (Eqs.~\ref{eq:eqs-of-motion-1} and \ref{eq:eqs-of-motion-2}). For low mass ratios, this assumption is valid, as the center of mass (CM) of the binary is close to the core of the giant star. For higher mass ratios, the CM is located in between the companion and the core of the giant star. Hence, both the core of the giant star and the companion are orbiting inside the CE around the CM and experience a drag because of dynamical friction. To some extent, this effect is compensated by using a free parameter for the drag-force calculations. For our 1D CE simulation with a mass ratio of 0.75, it is no longer valid to use this assumption, because the mass of the companion ($0.73\, \msun$) is larger than the mass of the helium core of the AGB star ($0.54 \, \msun$), \ie, the CM is located closer to the companion than to the core of the AGB star.

Additionally, the CE is simulated as the perturbed envelope of a giant single star. Similarly to the argument above, the CE is not centered on the core of the giant star, but rather on the CM of the binary. Therefore, the pressure, density, and sound speed at the location of the companion are different from the predictions in our model, which consequently causes the drag force to be different as well. Again, for low mass ratios, this assumption is valid, because the CM is located closer to the core of the giant star. However, this simplification breaks down for larger mass ratios.

In 3D simulations of CE events similar to the simulations in \citet{Sand2020}, spiral arms emerge as the companion plunges into the CE and the companion and the core of the giant star orbit each other inside the CE.\@ There are usually two spiral arms observed, one originating at the companion at the other originating at the core of the giant star. These spiral arms might transport orbital energy from the companion and the giant's core to the CE as they induce shocks in the envelope. This hypothesis needs further testing using 3D CE simulations. If true, an energy transport mechanism where the orbital energy is transferred to the envelope via shocks in the spiral arms would imply that the entire envelope could be heated in a 1D CE model. Additionally, there might be a time delay as the energy needs to be transported via the spiral arms first, before thermalizing in the envelope. Within our 1D model, we assume that all the released energy from the orbit is converted into heat. However, some of the orbital energy might also be converted into rotational energy, \ie, spinning up the envelope. We ignore this effect and only allow energy conversion into heat. \citet{Ivanova2016} argue that in 1D CE simulations, energy should not be added as heat but rather as kinetic energy. The spiral arms which are present in the simulations in \citet{Sand2020} convert some of the kinetic energy into thermal energy. Therefore, it seems reasonable to inject heat into the envelope to effectively simulate the response of the envelope to the spiral-in instead of kinetic energy.

During the early plunge-in phase, there is a difference in the envelope ejection mechanism between our 1D CE method and the 3D CE simulations in \citet{Sand2020}. In the 1D model, a large part of the envelope is slowly heated up. The injected internal energy is converted into kinetic energy and almost the entire envelope starts to expand. It takes a certain time until the surface of the envelope exceeds the escape velocity and becomes unbound. In the 3D simulations, the outer layers are dynamically flung out and become immediately unbound, \ie, orbital energy is directly converted to kinetic energy rather than heat. As our 1D model does not include energy injection as kinetic energy, the initial differences in the mass fractions of the ejected envelope are expected (see Figs.~\ref{fig:best_fit_q025} and \ref{fig:higher_q}). 

\vb{To account for the mass of the companion inside the envelope, we modified the gravitational constant outside the orbit of the companion (\sectref{sec:mod_G}). This is equivalent to modeling the companion as a thin shell with radius $a_\mathrm{orb}$, as such a thin shell has a constant potential for $r < a_\mathrm{orb}$ and a ${\sim}\, 1/r$ potential for $r > a_\mathrm{orb}$. When comparing this approximation to 3D CE simulations, \citet{Ivanova2016} find a deeper potential in 3D simulations close to the orbit and a more shallow potential far outside, with a relative deviation of up to $50 \, \%$ between the 1D potential with the thin shell approximation and the 3D simulations for a similar CE configuration. This approximation could be improved with a different potential form for the companion, but it works well to first order. Inconsistencies might also arise as we sometimes model the companion as a point particle, \eg, for evaluating the drag force and integrating the orbits (\sectref{sec:orbital_evolution}), and sometimes as a shell, \eg, for modeling the influence of the companion on the structure of the CE (\sectref{sec:mod_G}).}

\subsection{Advantages of the 1D CE model}\label{sec:dics-advantages}
The main reason for creating a 1D CE model is to reduce the computational cost of simulations, which enables larger parameter studies. Our 1D CE simulations take ${\sim}\, 10$ core-hours to simulate a physical time of \qty{2000}{d}, while the computational cost for the 3D simulation is of the order of $10^5$ core-hours. This means that the 1D CE simulation can be run easily on a desktop computer and does not rely on high-performance computing facilities. Therefore, a larger parameter space of systems that evolve through a CE phase can be explored with such 1D models.

In the simulation of \citet{Sand2020}, the core of the giant star is cut out and replaced by a point mass to allow for sufficiently large time-steps \citep{Ohlmann2016a, Sand2020}. The cut-out core of the giant star cannot respond to the dynamic changes in the envelope. In \citet{Sand2020}, the core is cut at $5 \, \%$ of the initial radius of the giant star, \ie, the cut-out core is larger than the helium core and contains hydrogen-rich layers. As the CE expands and is subsequently ejected, the core in the 3D simulation is expected to expand on the dynamical timescale to react to the loss of pressure by the CE. Our 1D simulations can resolve the core throughout the whole CE simulation and can capture the expansion of the layers that are considered to be inside the excised core in the 3D simulations. We find that the mass of the core, as defined in \citet{Sand2020}, decreases by ${\sim}\, 3 \, \%$ during our 1D CE simulations because the core expands. If the core is allowed to expand, we expect a deeper spiral-in as more envelope material needs to be ejected. This is a possible explanation for why we consistently find lower post-plunge-in separations compared to the 3D models \citet{Sand2020}.

In addition, the 1D simulations in \mesa include energy transport via \vb{photons} and photosphere cooling which is more difficult to implement 3D simulations. This also means that the release and transport of the recombination energy is better handled in the 1D model, which is another possible reason why we find lower post-plunge-in separations compared to the 3D simulations.

\subsection{Comparison to other 1D and 3D CE simulations}\label{sec:dics-comparison}

Several 1D CE models were proposed in the past to simulate CEs. The model of \citet{Fragos2019} is similar to our model in the sense that they assume a drag force acting on the companion, which removes orbital energy that is then injected as heat into the envelope. In contrast to our model, they assume a circular orbit for the companion and the released energy is injected into a region defined by one local accretion radius, \ie, heating within $a_\mathrm{orb} \pm R_\mathrm{a}(a_\mathrm{orb})$ using the terminology of our model. We find that the extent of the heating zone directly affects the envelope ejection and we need to tune the extent of the heated zone via \Ch to obtain the same amount of envelope ejection as in the 3D simulations. A similar model compared to \citet{Fragos2019} is used by \citet{OConnor2023} to study the envelope's response to a planetary engulfment. 
In our model, we have the advantage of comparing the predictions from the 1D simulation to 3D simulations of the same initial CE setup and tuning the free parameters of our model accordingly.

\citet{Sand2020} find in their 3D CE simulations that the energy from recombination is necessary to eject significant amounts of the envelope during the simulation by comparing different equations of state, which either include or do not include recombination energy. In our 1D simulations, we also find that the envelope ejection is only triggered once hydrogen recombines. A significant fraction of the recombination energy accelerates the envelope, which is subsequently ejected. Once recombination starts, the hydrogen recombination front stays at a constant radius ($\qty{700}{d} \leq t \leq \qty{1300}{d}$ in \figref{fig:kipp1}b). This process is similar to the \emph{steady recombination outflow} described by \citet{Ivanova2016}. The recombination energy expands and ejects the envelope. Therefore, the inner layers adjust to the decrease in pressure by expanding. As these inner layers expand, they cool down until recombination expands them further, causing the even deeper layers to expand, and so on. \citet{Ivanova2016} argue that this leads to the recombination at a constant radial coordinate. In the late stages of the simulations, we find, however, that the radius of the recombination front of hydrogen and also helium decreases. This might be caused by the envelope running out of material, since more than $97 \, \%$ of the initial envelope recombined at the end of the simulations.

\subsection{Physical motivation for two free parameters}\label{sec:disc-two_parameters}
We use the drag force models of \citet{Kim2010} and \citet{Kim2007} for all of our 1D CE simulations. Both models assume a perturber moving on a circular orbit through a gaseous background medium. While \citet{Kim2007} assume a low-mass perturber, the model of \citet{Kim2010} applies also to high-mass perturbers. In both cases, the background medium is assumed to be homogeneous in density. \citet{Kim2010} argue that density gradients can be ignored if the Bondi radius is much smaller than the pressure scale height, \ie, $\mathcal{M} \ll (M_{1,a_\mathrm{orb}} / M_2)^{1/4}$. In our 1D CE simulations, we find that this condition is not always fulfilled, especially during the plunge-in phase. Additionally, \citet{Kim2010} assumes that the centrifugal force and the Coriolis force can be ignored. This assumption is valid if $\mathcal{B} / \mathcal{M}^2 \ll 1$. We also find that this condition is not satisfied during our simulations. Because our simulations disagree with these assumptions, we do not expect the drag force given by the models of \citet{Kim2010} and \citet{Kim2007} to provide the correct values for our simulation without considering a calibration factor. Hence, it seems appropriate to introduce a calibration factor \Cd for the drag force, which we adjust by comparing our 1D to the 3D CE simulations. From the three comparisons between 1D and 3D simulations, we find that $C_d \approx 0.25$ (\tabref{tab:summary}). 

The Bondi-\vb{Lyttleton}-Hoyle model for accretion \citep{Hoyle1939,Bondi1944} also assumes a homogeneous background density. We use the accretion radius of the Bondi-\vb{Lyttleton}-Hoyle model to estimate the heating zone, \ie, the layers where the material is gravitationally deflected and focused by the companion. Because we assume spherical symmetry in the 1D CE model, we always heat a spherical shell defined by the accretion radius. In a real CE scenario, only a region around the companion might be gravitationally affected, and not a spherical shell centered on the core of the giant star. Introducing a second calibration factor \Ch for the accretion radius and therefore also for the size of the heating zone seems reasonable. From the results of our simulations, we find the trend of decreasing \Ch for increasing $q$ (\tabref{tab:summary}). We also analyzed the quantities $q R_\mathrm{a}(a_\mathrm{orb}) / a_\mathrm{orb}$ and $q (r^\mathrm{heat}_\mathrm{max} - r^\mathrm{heat}_\mathrm{min}) / a_\mathrm{orb}$ to see if we can find a calibration for the extent of the heating zone. For the three 1D models we computed, we could not see a better calibration with these quantities. Therefore, we kept using the heating parameter \Ch as a free parameter in our model.

From the simulations discussed in \sectref{sec:res_varying_Cd_Ch}, it is necessary to have two free parameters in our 1D CE model. If we restrict our model to only one free parameter, we cannot reproduce the results of the 3D CE simulations of \citet{Sand2020}. \vbre{Many approximations of our 1D model are captured by the free parameters \Cd and \Ch. Therefore, the best-fitting values do not directly carry any physical meaning and cannot be immediately used to constrain the physics during CE events.}

%
%
\section{Conclusions}\label{sec:conclusion}
We presented a 1D approach to simulate the CE evolution within the stellar evolution code \mesa using its hydrodynamic capabilities. The CE is modeled by the envelope of a giant star, in which a point-mass companion is placed on a circular orbit. The viscous forces hinder the motion of the companion because of the dynamical drag by the CE. These extra forces are included in the equations of motion of the companion via a parametric drag-force prescription, allowing us to integrate the orbital evolution of the binary star. The energy lost because of the drag force is added as heat to the envelope. We trace the layers expanding faster than the escape velocity and remove them from the simulated domain. To fit our results to the 3D CE simulations, we include two free parameters, which we use as optimization parameters for the fits. 

We simulate the CE phase of a \qty{0.97}{\msun} AGB star and a point-mass companion with mass ratios $q = 0.25$, 0.50, and 0.75. When comparing and fitting these simulations to the 3D CE simulations of \citet{Sand2020}, we find the following conclusions:
\begin{itemize}
    \item It is possible to reproduce the spiral-in curve and the mass fraction of the ejected envelope when using both free parameters for $q = 0.25$ and 0.50. With one parameter alone, this is not possible. Therefore, 1D CE models similar to ours probably require at least two free parameters to reproduce 3D CE computations. We find that the spiral-in timescale is mostly determined by \Cd, while the mass-fraction of the ejected envelope is determined by both \Cd and \Ch.
    \item We are unable to find satisfactory fits with the 1D CE model for $q = 0.75$. This might be an extreme case where the assumptions and approximations of our 1D CE model are no longer valid and do not result in a physical solution.
    \item For all mass ratios, we find the post-plunge-in separation in the best-fitting 1D simulation to be smaller than those of the 3D simulation.
    \item Regardless of \Cd and \Ch, the post-plunge-in separation is almost always similar and only depends on the mass ratio. This suggests a deeper physical mechanism that determines the post-plunge-in separation. 
    \item The recombination energy released from hydrogen and helium likely plays an important role in accelerating and ejecting the envelope.
\end{itemize}

In a future study, we plan to simulate CE events with different initial configurations, \ie, different masses, and evolutionary stages of the giant star, and compare them to 3D CE simulations to investigate whether a global calibration of the two free parameters based on the mass and evolutionary stage of the giant star as well as the mass ratio is possible. If we can find such a calibration, this CE model can be used to predict the outcome of CE events, \ie, the post-plunge-in orbital separations and the ejecta mass if the envelope is ejected successfully.

\begin{acknowledgements}
    \vb{We thank the anonymous referee for their feedback, which helped to improve the quality of the paper.} We thank C. Sand for providing the data for \vb{the} simulations described in \citet{Sand2020}.

    VAB, FRNS, PhP and FKR acknowledge support from the Klaus Tschira Foundation.
    This work has received funding from the European Research Council (ERC) under the European Union’s Horizon 2020 research and innovation programme (Grant agreement No.\ 945806). This work is supported by the Deutsche Forschungsgemeinschaft (DFG, German Research Foundation) under Germany’s Excellence Strategy EXC 2181/1-390900948 (the Heidelberg STRUCTURES Excellence Cluster). VAB acknowledges support from the International Max Planck Research School for Astronomy and Cosmic Physics at the University of Heidelberg (IMPRS-HD).
\end{acknowledgements}

\bibliographystyle{aa}
\bibliography{references_for_1D_CE}

\begin{appendix}
\section{Resolution study}\label{apdx:resolution_study}
\begin{figure}
    \resizebox{\hsize}{!}{\includegraphics{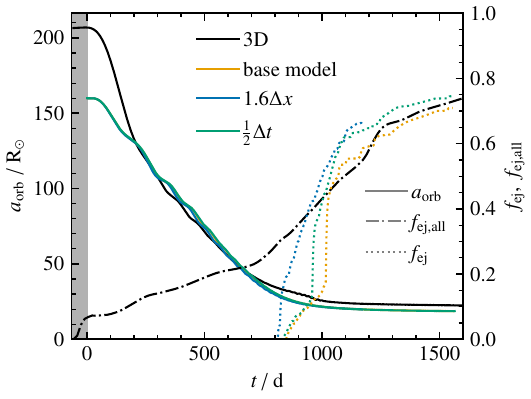}}
    \caption{\vb{Similar to \figref{fig:best_fit_q025}, but also showing the simulation at a lower spatial resolution and at a higher time resolution.}}
    \label{fig:resolution}
\end{figure}

\vb{To prove the robustness of the results we obtained with our 1D CE method described in \sectref{sec:methods}, we perform a small resolution study. We rerun the simulations for $q=0.25$, presented in \sectref{sec:results_q0.25}. For one simulation, we increase the time resolution by a factor $2$, for a second simulation we decrease the spatial resolution by a factor of $1.6$. The results of these simulations are shown in \figref{fig:resolution}.}

\vb{The spiral in the behavior of the companion is mostly unchanged when compared to the original simulation. There are however some differences in the envelope ejection between the different simulations. The exact timing of the main envelope-ejection event varies by about \qty{100}{d}. In addition, the shape of the ejection curve also depends on the resolution. Once a more continuous ejection is established after the main ejection events at $t = 900 - 1000 \, \mathrm{d}$, the values of $f_\mathrm{ej}$ only vary by \qty{5}{\%}, \ie, within our expected tolerances.}

\vb{The reason for the differences in the envelope ejection originates mostly from differences in the surface properties between the simulations. As described in \sectref{sec:relaxation}, we perform a relaxation run before the CE simulations where we switch the outer boundary condition and turn on the hydro mode. These changes cause small amplitude oscillations close to the surface of the CE (see also \figref{fig:kipp1}a). The oscillations are stochastic and therefore also some of the surface properties, \eg, surface velocity. This means that the exact timing of when the surface velocity exceeds the escape velocity has a minor dependence on the initial oscillations. This causes the difference in the starting time of the ejection. The total ejected mass towards the end of the simulations, the quantity in which we are mostly interested varies within our tolerances.}

\vb{Based on this resolution study, we conclude that the final values for the orbital separation do not depend on the resolution. The total ejecta mass has a minor dependence on the resolution but varies within our tolerances of a few percent.}

\end{appendix}

\end{document}